# Antiferroelectric Order in Nematic Liquids: Flexoelectricity Versus Electrostatics

*Peter Medle Rupnik, Ema Hanžel, Matija Lovšin, Natan Osterman, Calum Jordan Gibb, Richard J. Mandle, Nerea Sebastián, and Alenka Mertelj\**

The recent discovery of ferroelectric nematic liquid crystalline phases marks a major breakthrough in soft matter research. An intermediate phase, often observed between the nonpolar and the ferroelectric nematic phase, shows a distinct antiferroelectric response to electric fields. However, its structure and formation mechanisms remain debated, with flexoelectric and electrostatics effects proposed as competing mechanisms. By controlling the magnitude of electrostatic forces through ion addition in two representative ferroelectric nematic materials, it is shown that the primary mechanism for the emergence of antiferroelectric order is the flexoelectric coupling between electric polarization and splay deformation of the nematic director. The addition of ions significantly expands the temperature range over which the antiferroelectric phase is observed, with this range increasing with increasing ion concentration. Polarizing optical microscopy studies and second harmonic generation (SHG) microscopy reveal the splayed structure modulated in 2D, while SHG interferometry confirms its antiferroelectric character. The model previously used to describe pretransitional behavior is extended by incorporating the electrostatic contribution of ions. The model shows qualitative agreement with the experiments, accurately reproducing the phase diagram and temperature-dependent evolution of the modulation period of the observed structure.

## 1. Introduction

The discovery of ferroelectric nematic liquids in 2017[1–3] was followed by numerous studies[4] of their properties and the discovery of other new phases in these materials such as helielectric,[5–7] polar twist bend nematic phases,[8,9] and a variety of new ferroelectric smectic phases.[9–16] Besides being interesting from the fundamental point of view they are very promising as switchable ferroelectric materials, e.g., in nonlinear optics and as sources of quantum light.[17] Both materials reported in 2017, RM734[2] and DIO,[1] in addition to the usual nematic and ferroelectric nematic phases, also exhibit an antiferroelectric nematic phase, which appears in between both phases. In DIO, this intermediate phase, which can be polarized by an external electric field, has been recognized as an individual phase from the beginning.[1] In contrast, in RM734 its existence was inferred from polarized optical microscopy (POM) observations[4,18,19] but was only lately confirmed also by precision calorimetry[20] in a short temperature range. The appearance of the intermediate phase between the nematic and ferroelectric nematic phases seems to be a common feature of many materials.[12,21–25]

A characteristic double-peak response of the current to an external triangular voltage suggests that the intermediate phase is antiferroelectric.[26,27] However, the structure of the phase and the driving mechanism for its appearance are still under debate, as reflected by the range of nomenclatures found in the literature to refer to it, i.e., $M_2$, $N_x$, $N_S$, and $SmZ_A$.[4] In solids, the transition to the antiferroelectric phase is a structural transition during which two sublattices having antiparallel polarization form.[28] The phase is nonpolar and is distinguished from other nonpolar phases by exhibiting a double hysteresis.[28,29] While the driving forces for the appearance of the ferroelectric order in solids are well understood,[30] the reason why some materials also exhibit an antiferroelectric phase is less clear. The flexoelectric coupling has been shown to be a mechanism driving the transition from the precursor paraelectric to the antiferroelectric phase in some cases,[28,31] however, whether the role of flexoelectricity is universal for all solid antiferroelectrics is not clear.[28]

In a nematic liquid in contrast to a crystal, there is no underlying periodic structure so the mechanisms leading to the polar, i.e., ferroelectric and antiferroelectric order are different than in

P. Medle Rupnik, E. Hanžel, M. Lovšin, N. Osterman, N. Sebastián, A. Mertelj
Jožef Stefan Institute
Ljubljana 1000, Slovenia
E-mail: alenka.mertelj@ijs.si
P. Medle Rupnik, E. Hanžel, M. Lovšin, N. Osterman
Faculty of Mathematics and Physics
University of Ljubljana
Ljubljana 1000, Slovenia
C. J. Gibb, R. J. Mandle
School of Chemistry
University of Leeds
Leeds LS2 9JT, UK
C. J. Gibb, R. J. Mandle
School of Physics and Astronomy
University of Leeds
Leeds LS2 9JT, UK

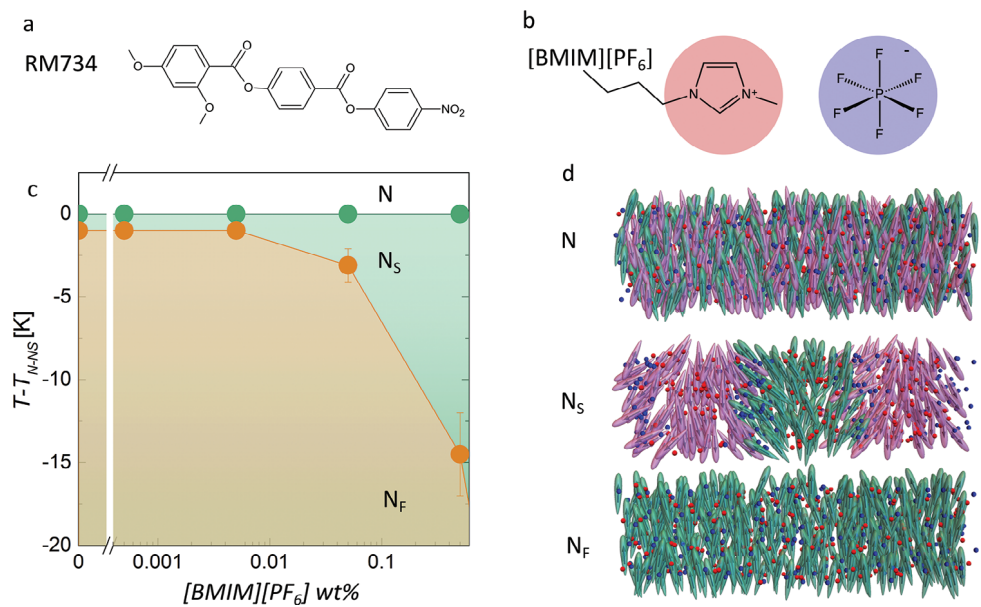

**Figure 1.** a) Molecular structure of RM734. b) Chemical structure of the ionic liquid [BMIM][PF$_6$] (Sigma–Aldrich). c) Phase diagram of the RM734 + [BMIM][PF$_6$] mixture, for [BMIM][PF$_6$] wt.% (mixtures with 0, 0.0005, 0.005, 0.05 and 0.5 wt.% are considered). d) Schematic representation of the different phases of the mixtures; nematic (N), splay nematic (N$_S$), and ferroelectric nematic (N$_F$).

solids. The studies of the pretransitional behavior at the transition from the paraelectric to the antiferroelectric nematic phase in RM734[18,19] demonstrated that flexoelectric coupling plays a major role. Upon cooling toward the transition, simultaneously strong softening of splay orientational elastic constant and divergent behavior of electric susceptibility is observed that shows that the transition is a ferroelectric-ferroelastic phase transition.[19] This behavior can be well explained by the model in which polarization and splay deformation are coupled via flexoelectric coupling.[18,19,32] This model predicts the decrease of the effective splay elastic constant with increasing electric susceptibility, and when the elastic constant reaches zero, the uniform nematic phase becomes unstable toward splay deformation. Because it is not possible to fill the space with uniform splay, the structure can become periodic either in 1D[18] or 2D[32] with simultaneously alternating splay deformation and the sign of electric polarization.

Although the flexoelectric model captures the physics of the pretransitional behavior well in RM734, an alternative explanation has been offered for the material DIO.[27] From the small angle X-ray scattering (SAXS) measurement that showed a weak peak corresponding to a distance of 8.8 nm, the resonant X-ray scattering, which is sensitive to variations in the orientation of the molecules, that showed a peak corresponding to twice the distance obtained by SAXS, and the POM observation of zig-zag defects, the authors concluded that the phase is an antiferroelectric smectic phase with the modulation perpendicular to the director and named it SmZ$_A$. In the model used to describe the phase, the flexoelectric coupling is neglected with the argument that the depolarization field, that arises due to the splay deformation, suppresses the splay, and the driving mechanism for the formation of the SmZ$_A$ phase are electrostatic forces. The studies of the mixtures of RM734 and DIO, however, showed that the temperature region of the intermediate phase grows continuously with increasing DIO concentration suggesting that in both materials the intermediate phase is the same.[33,34]

By precisely tuning the local electric field, we can explore such competition between flexoelectric coupling and electrostatics. Here, we demonstrate that the flexoelectric coupling is the primary driving mechanism behind the emergence of the antiferroelectric nematic phase. This is achieved by controllably adjusting the concentration of free ions in ferroelectric nematic materials, and with it reducing the local electric field. For this, the paradigmatic ferroelectric nematic materials RM734 (**Figure 1**a) and FNLC-1571 (Merck Electronics KGaA) were mixed with a small amount (from 0.0005 to 5 wt.%) of ionic liquid [BMIM][PF$_6$] (Sigma–Aldrich) (Figure 1b), to achieve ion number densities $\rho_N$ ranging from $10^{22}$ to $10^{26}$ ions m$^{-3}$. We show that increasing the ion concentration promotes the antiferroelectric phase, extending its temperature range and revealing an SHG-active 2D-modulated antiferroelectric splayed structure, to which hereinafter we will refer to as N$_S$ phase.

## 2. Results

### 2.1. N$_S$ Phase in Mixtures of RM734 and Ionic Liquid

The material RM734 exhibits a transition from the nematic (N) to the antiferroelectric nematic phase (N$_S$), which has a temperature range of ≈1 K, followed by the transition to the ferroelectric nematic (N$_F$) phase.[4,20] During cooling, when approaching the transition from N to N$_s$ phase, the splay fluctuations exhibit critical behavior, i.e., their amplitude strongly increases while their relaxation rate strongly decreases. This is accompanied by the growth of the polar order.[19] At the phase transition to the N$_s$ phase, the splay fluctuations freeze, which in POM is seen as the absence of strong flickering observed in the precursor nematic phase. Just before the transition to the N$_F$ phase, a stripy pattern is observed.[4,18] In the N$_F$ phase, fluctuations again become stronger that in POM look like flickering, similar to that observed in the N phase away from the phase transition temperature. In the mixtures of RM734 and ionic liquid [BMIM][PF$_6$], the temperature range of the N$_s$ becomes wider (Figure 1c). This widening becomes already visible in POM at [BMIM][PF$_6$] concentrations of ≈0.005 wt.% (which corresponds to $\rho_N \approx 10^{23}$ ions m$^{-3}$). At higher concentrations (≈0.5 wt.%), the N to N$_s$ phase lowers 2–3 K and the temperature width of the antiferroelectric phase becomes of the order of 15 K (Figure 1c; Figure S1, Supporting Information). The width is very sensitive to the amount of added ionic liquid and variations are observed, however, the described features are the same for all cases. All samples become visible in SHG-M at the N–N$_s$ phase transition (Figure S2, Supporting Information). The average intensity of SHG microscopy (SHG-M) images increases during the transition, then it decreases and almost disappears. It becomes strong again right before the N$_s$–N$_F$ transition (Figure S2, Supporting Information).

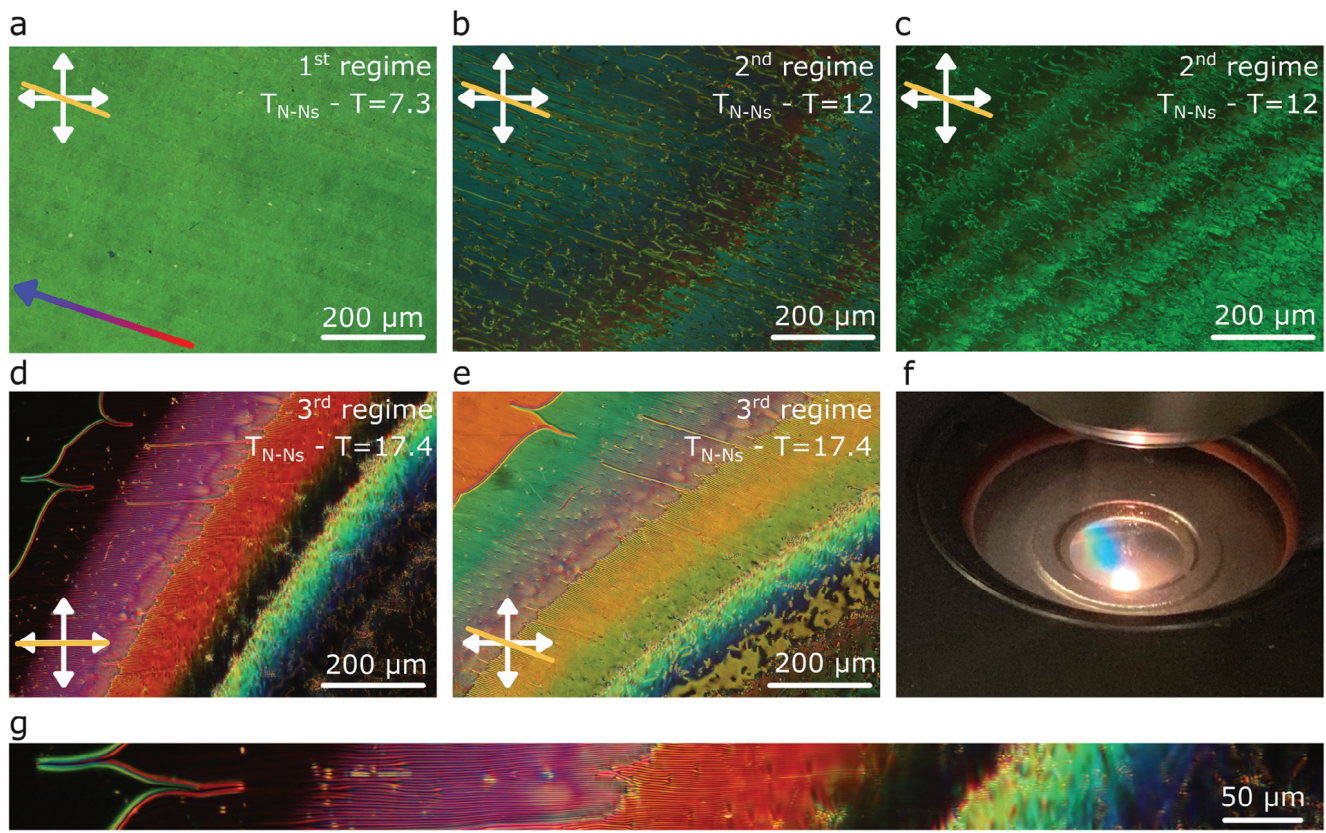

**Figure 2.** Polarizing optical microscopy images of the different regimes observed in the NS phase of RM734 + 0.5 wt.% [BMIM][PF$_6$] in a 5 µm thick cell with parallel rubbing during cooling and under a small temperature gradient. a) POM image in the first regime. b) Texture in the second regime, which is characterized by the appearance of defects and birefringence jumps. c) Texture in the second regime as observed with monochromatic light ($\lambda$ = 546 nm) d,e) Microphotografs in the third regime showing distinct modulated texture with increased modulation period on cooling. f) Photograph of the LC cell displaying Bragg scattering in the temperature range of the 3rd regime. g) Enlarged region from (d). Double-headed arrows indicate the direction of the crossed polarizers. The yellow line indicates the LC cell's rubbing direction. The gradient-colored arrow indicates the direction of the temperature gradient, with the blue color (arrowhead) indicating the lower temperature end.

We focus here on the mixture with 0.5 wt.% of [BMIM][PF$_6$] ($\rho_N \approx 10^{25}$ ions m$^{-3}$). In this system, the POM observations of the intermediate antiferroelectric nematic phase can be divided into three temperature regimes. Distinction between the regimes becomes even more evident in a slight temperature gradient in the confining LC cell (<0.25 K mm$^{-1}$, see Experimental Section). In the temperature region of ≈9 K just below the N–N$_s$ phase transition, the POM images of the sample strongly resemble those observed in the antiferroelectric modulated phase (**Figure 2a**) as for example in DIO or FNLC-1571.[4] The second regime is distinguished by the appearance of more defects, and by several jumps in the effective birefringence $\Delta n_{eff}$, which abruptly decreases for ≈0.05. After a jump $\Delta n_{eff}$ continuously increases, then it again abruptly decreases, and so on (Figure 2b,c). In 5 µm thick cells with parallel rubbing, 13–15 such jumps are observed in the temperature range of ≈5 K. Last, in the third regime, a periodic pattern becomes visible with stripes along the rubbing direction (**Figures 2d,g** and **3**). This pattern causes Bragg scattering that is observed by the naked eye (Figure 2f). In 5 µm thick cells, the stripes are observed in a temperature range of ≈0.5 K, in which three regions can be distinguished (Figure 3a,b). Within each region, the modulation period increases with cooling and then abruptly reduces at the beginning of the next region (Figure 3c). These abrupt jumps in the structure correspond to the chaining of defects. Finally, the periodic pattern smoothly fades and the sample appears uniform. Upon further cooling, two defect lines materialize separating this homogenous region from another homogeneous region. In cells with parallel rubbing, both homogeneous regions show good extinction between crossed polarizers, but exhibit different colors when the sample is rotated (Figure 2d,e), which indicates different effective birefringence. This can be explained by the partial out-of-plane orientation of **n** due to the splayed structure (Figure S3b, Supporting Information). The two defect lines dividing the homogenous regions are preferably located together (Figure 2d). If they are separated, for example, due to the surface imperfections, a region of twisted structure emerges between them (Figure S3a, Supporting Information).

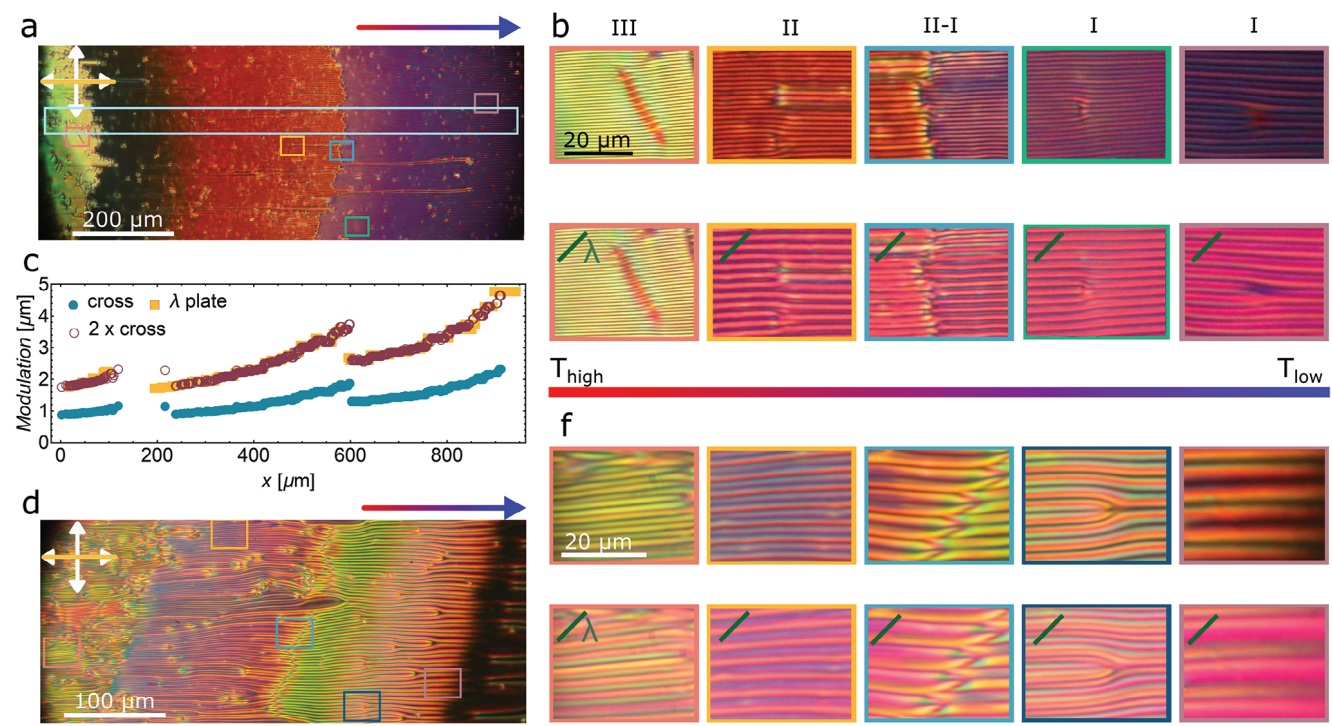

**Figure 3.** Comparison of periodic textures observed for RM734 + 0.5 wt.% [BMIM][PF$_6$] in 5 and 10 μm thick cells with parallel rubbing under a small temperature gradient (0.25 K mm$^{-1}$). a) Overview POM image for $d = 5$ μm. b) Zoom-in images from the highlighted areas in (a) as observed under crossed polarizers (top row) and with a $\lambda$ plate inserted at 45 degrees (bottom row). c) Modulation period (full symbols) along the elongated light-blue area highlighted in (a). Empty symbols correspond to 2 times the modulation obtained under crossed polarizers. d) Overview POM image for $d = 10$ μm. f) Zoom-in images from the highlighted areas in (d) as observed under crossed polarizers (top row) and with a $\lambda$ plate inserted at 45 degrees (bottom row). The gradient-colored arrows indicates the direction of the temperature gradient, with the blue color (arrowhead) indicating the lower temperature end.

Similar characteristics can be observed upon heating from the second (i.e., the low temperature) homogeneous structure. First, the stripes nucleate and grow from imperfections. Additionally, regions with 2D modulated structures have also been observed (Figure S4, Supporting Information). On further heating, the stripes become thinner and many defects appear. Eventually, the structure becomes smoother, however, less homogenous than that observed during cooling in the first regime.

To explore how the thickness of the LC cell affects the observed periodic structures in the third temperature regime we compared 5 and 10 μm thick cells with parallel rubbing (**Figure 3**). While the maximal value of the modulation period observed in 5 μm cells is ≈5 μm, in the 10 μm cell, the maximal modulation period is ≈10 μm. Moreover, instead of the 3 distinctive regions observed for 5 μm cells, in the thicker case, there are 5 clearly discernable regions where stripes are visible (Figure S5, Supporting Information). The analysis of POM images between crossed polarizers, both with and without additional $\lambda$ plate, evidence that the modulation period is twice that observed in the extinction position. From the inspection of the optical behavior under different conditions (i.e., sample rotation or slight uncrossed analyzer, Figure S6, Supporting Information) the structure can be inferred. Within each period there are stripes with opposite twist deformation along the thickness of the cell. That is, at the surface, the orientation is uniform as prescribed by the rubbing, away from the surface it exhibits in-plane splay modulation (Figure 1d) that results in two regions with opposite twist deformation. On cooling, the increase of the modulation period causes the appearance of edge dislocations (Figure 3b–f, for example in region I), which are present in all regions where stripes are visible.

In RM734, the SHG signal is the strongest when the incoming laser beam is polarized along **n** (Methods).[35] The dependence of SHG-M intensity on the pump laser polarization in the periodic structure in the 10 μm thick cell shows that the angle of incoming polarization at which the SHG intensity is the strongest Φ varies periodically from ≈−20 deg to 20 deg (**Figure** 4b,c). As in the direction perpendicular to the layer the splay angle varies and the structure is inhomogeneous on the scale of a few microns the diffraction of the pump laser and SHG beam generated in the sample are different, which makes the interpretation of the SHG images more difficult. This is the reason why the angle Φ, while giving the information on the average orientation of **n** and the order of magnitude of the splay amplitude $\theta$, cannot be used to determine the exact value of $\theta$. However, the SHG-M confirms the POM observations, according to which the director exhibits a modulated splayed structure.

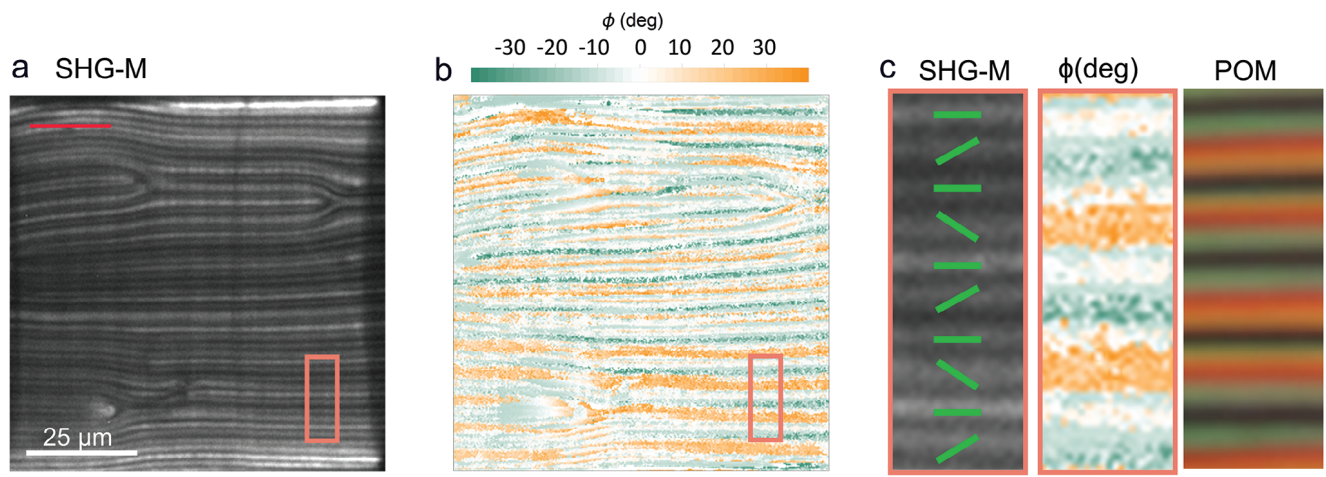

**Figure 4.** Second Harmonic Generation microscopy (SHG-M) for RM734 + 0.5 wt.% [BMIM][PF$_6$] in 10 µm cell with parallel rubbing in the last periodic structure (I). a) SHG-M microphotograph for incoming pump polarization along the lines as indicated by the red bar. b) Analysis of the maximum SHG intensity as a function of the direction of the incoming pump polarization showing the angle at which maximum intensity is observed (0 degrees correspond to horizontal direction). c) Zoom-in image of the highlighted area in (a&b) comparing SHG microphotograph, angle for maximum intensity analysis, and POM image. Green lines show the average direction of **n** as deduced from laser polarization giving the strongest SHG intensity.

The addition of 5 wt.% of [BMIM][PF$_6$] causes the decrease of the N–N$_s$ transition temperature for ≈22 K in comparison to pure RM734 (Figure S7, Supporting Information). About 26 K below the phase transition temperature, the material under POM appears to have a grainy texture, which becomes more pronounced with further cooling, and at ≈35 K below N–N$_S$ transition the sample exhibits transition to a phase that is optically isotropic. The sample becomes weakly visible in SHG-M at the N–N$_S$ phase transition and remains so until a grainy structure appears. There the intensity of the M-SHG images becomes stronger, exhibits a peak, then with further cooling decreases, and when the material enters the optically isotropic phase, it completely disappears (Figure S7, Supporting Information). This optically isotropic phase has been studied in ref. [36].

### 2.2. N$_S$ Phase at Room Temperature in FNLC-1571 Mixture with Ionic Liquid

To investigate the effect of ion addition in materials with a wider antiferroelectric nematic phase, we focused on the mixture FNLC-1571 (Merck Electronics KGaA), which exhibits the N$_S$ phase in a temperature region of ≈14 K.[37] In POM, the transition from antiferroelectric N$_S$ to N$_F$ phase happens by the appearance of a striped texture, which in the material without added ions, quickly smoothens. The addition of 0.6 wt.% of [BMIM][PF$_6$] broadens the temperature range over which the antiferroelectric phase is observed, stabilizing the striped texture at room temperature with a well-defined periodicity of ≈3 µm (**Figure** 5b), which results in strong Bragg scattering (Figure 5h).

This observation proves that N$_S$ phase is thermodynamically stable and not merely the result of thermal gradients during the transition. In the 10 µm thick cell with parallel rubbing (Figure 5f,g), the structure is similar to that in the 5 µm thick cell (Figure 5a–e) and has a similar modulation period. Besides regions with the stripes oriented along rubbing, there are regions where the stripes' direction changes along the thickness of the cell showing that indeed the structure consists of stripes and not layers (Figure 5g). To assess the influence of the surface anchoring on the periodic structure we compared the cell with parallel rubbing with a cell coated with hydrophilic molecules, i.e., nonafluorohexyltriethoxysilane (Figure S8, Supporting Information). In this case, randomly oriented regions with modulated structure were observed in all parts of the sample. At the edge, a region ≈60 µm wide developed with better-oriented stripes (Figure S8, Supporting Information). In the cells with ITO electrodes on both sides and without any additional layer, we observed regions with nicely oriented modulated structure separated by a chain of defects (**Figure** 6a). Remarkably, outside the cell's ITO area, the modulated structure forms a nematic-like superstructure with a rich variety of topological defects (Figure 6b; Figure S9, Supporting Information). Some of them resemble those typically observed in the Schlieren texture with topological charges ±1/2, ±1, however, there are also defects with a topological charge of 3/2 (Figure 6b). It appears that the superstructure can easily adapt by bending and splaying. Heating to the nematic phase results in the relaxation of most of the defects, on cooling they appear again with similar but not entirely equal structures indicating partial surface memory (Figure S10, Supporting Information). In the N phase, the defects in the Schlieren texture are typically stabilized by surface irregularities. The differences in the observed structures suggest that, in the ITO areas, there are fewer surface irregularities or that these irregularities have less impact compared to the regions outside the ITO areas. Furthermore, the two areas also differ in their electrostatic boundary conditions.

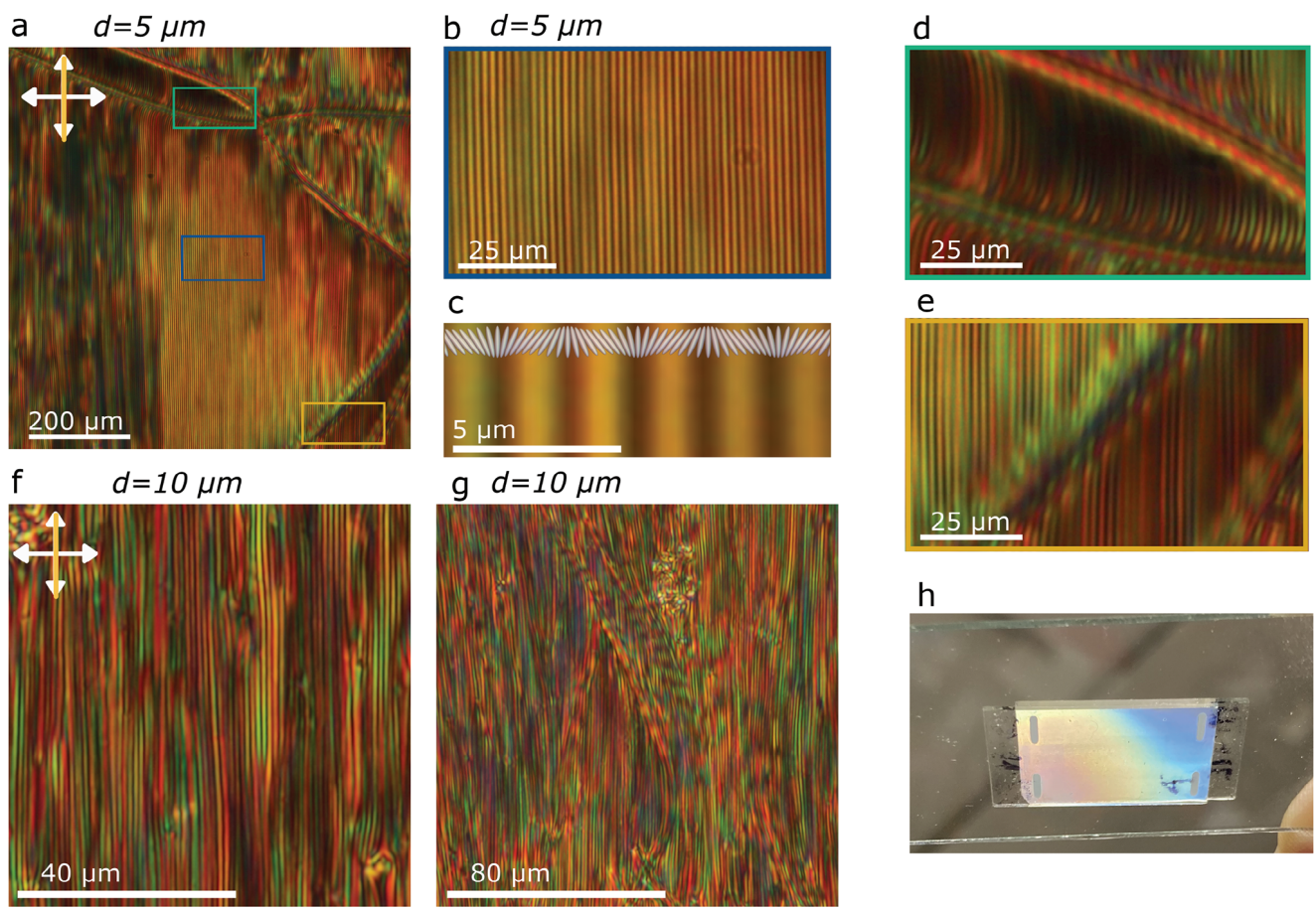

**Figure 5.** Modulated structure observed for FNLC-1571 + 0.6 wt.% [BMIM][PF$_6$]. a–e) Textures in 5 μm parallel rubbed cells. b–e) Zoom-in images from the highlighted areas in (a). f,g) Textures observed in 10 μm parallel rubbed cells. h) Photograph of FNLC-1571 + 0.6 wt.% [BMIM][PF$_6$] filled into a 5 μm parallel rubbed cells at room temperature showing strong Bragg scattering.

Both, the pure material and that with 0.6 wt.% of [BMIM][PF$_6$] become visible in M-SHG when a stripy texture appears and remain SHG active when the sample is cooled to room temperature. In the sample with 0.6 wt.% of [BMIM][PF$_6$] in the 5 μm thick cell, the stripes corresponding to the modulated structure observed in POM are well distinguished. The SHG-I microscopy was used to probe whether the polar order in neighboring stripes is different. As in the case of RM734, the polar orientation changes sign within the cell thickness. The inhomogeneous structure on the scale of a few microns causes different diffraction of the pump laser, SHG reference beam, and sample SHG beam. Consequently, the interference between the reference and sample SHG is suppressed and difficult to interpret. In some parts, the interference signal is strong enough, showing that within one modulation period, there are two regions with opposite phases (**Figure 7**) and thus proving the antiferroelectric character of the structure.

## 3. Discussion

Experimentally, a strong influence of flexoelectric coupling on the pretransitional behavior has been observed in the paraelectric nematic phase[18,19] and exploited in the ferroelectric nematic phase.[35] The flexoelectric effect is a consequence of the molecular shape.[38] In the apolar nematic phase, the molecules are on average oriented along the director **n**, with no distinction between **n** and –**n**. Molecules usually lack head–tail symmetry, i.e., they are of a wedge or a pear shape, so when there is a splay deformation from the packing point of view it's energetically favorable that more molecules are oriented in one direction (**Figure 8a**). If additionally, they have a nonzero component of a dipole moment along the long axis, such a splayed structure will have electric polarization (Figure 8a). In the polar nematic phase, most molecules are already oriented in the same direction, so from the entropic point of view spontaneous splayed structure is energetically favorable (Figure 8a). Since a homogeneous splay deformation cannot fill the space, possible solutions for the resulting structures are modulated antiferroelectric splay structures periodic in either 1D or 2D (Figure 8b,c).[18,32]

As was recognized already by Frank,[39] if the polarization is parallel to the average orientation of the molecules, spontaneous splay, characterized by a nonzero value of $\nabla \cdot \mathbf{n}$, results in divergence of the polarization, which is a source of the depolarization field. This depolarization field forces the polarization to unsplay. Consequently, the polar nematic phase is intrinsically frustrated – from the packing, i.e., entropic point of view, spontaneous splay is favorable, while the electrostatics disfavor splay deformation. So, when the amplitude of the polarization is small, the tendency to splay prevails and the phase is splayed and antiferroelectric. With increasing polar order, the electrostatic forces start to prevail over the tendency to splay, and when there is no splay deformation left there is no reason for the system to exhibit modulated antiferroelectric structure, so it undergoes the transition to the ferroelectric nematic phase. By adding free ions to the system, the depolarization field is partially screened and the tendency to spontaneous splay should become more pronounced, which should result in enhanced stability of the antiferroelectric splayed phase. And this is indeed what happens in the systems studied in this paper. The addition of ionic liquid causes the widening of the antiferroelectric N$_S$ phase in RM734 and FNLC-1571. In both cases, the modulation period increases upon cooling up to a value of several micrometers, becoming visible optically. When it reaches the thickness of the LC cell it continuously transforms into a homogenous structure. In both materials, the POM observations suggest that the structure has 2D splay modulation that supports the idea that the flexoelectric coupling is the mechanism leading to the formation of the antiferroelectric nematic phase.

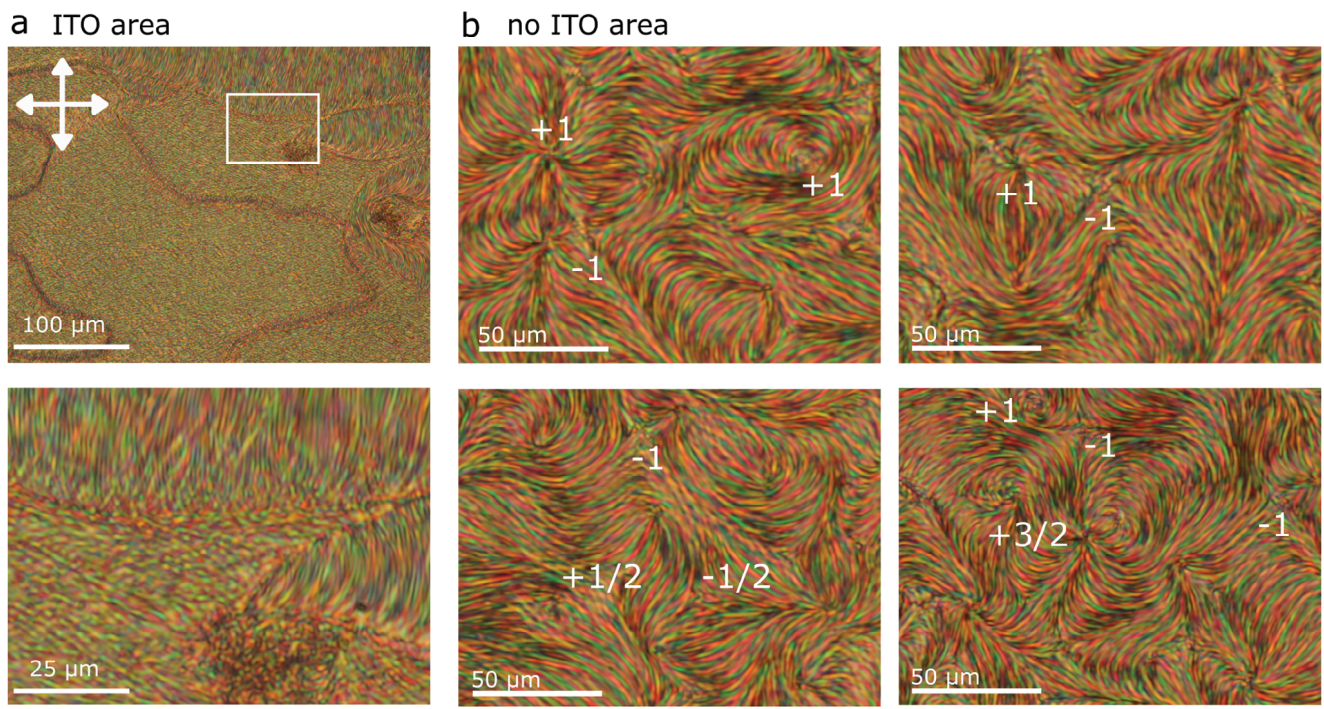

**Figure 6.** Polarizing optical microscopy textures observed for FNLC-1571 + 0.6 wt.% of [BMIM][PF$_6$] in a 10 μm LC cell without aligning layers at room temperature. a) Textures observed in the cell region with ITO electrode. b) Details of the superstructure formed in the region outside the electrodes show a rich variety of topological defects. Areas in b correspond to highlighted regions in Figure S9 (Supporting Information). Images are taken between crossed polarizers, in the vertical/horizontal direction.

The single and double splay antiferroelectric nematic phases (Figure 8b,c) are characterized by the modulation period $\Lambda$ and the amplitude of the splay modulation $\theta$ defined as the maximal angle for which the director **n** deviates from the orientation of the stripes, and the splay curvature $1/R$ (Figure 8d). In RM734 with 0.5 wt.% [BMIM][PF$_6$] the observed jumps in birefringence can be attributed to the steps in the number of modulation periods that fit within the thickness of the cell – an effect similar to Grandjean-Cano steps in cholesteric liquid crystal in a wedge cell. A possible explanation is that the splay curvature decreases with decreasing temperature in either a 1D or 2D splay nematic phase (Figure 8f,g). If the ideal splay curvature decreases, the modulated structure can adapt in two ways, either i) the modulation period $\Lambda$ increases or ii) the amplitude of the splay deformation decreases (Figure 8d). The first scenario is energetically more favorable because the volume with ideal splay curvature increases and, simultaneously, there is a smaller number of energetically unfavorable regions in which the polarization changes sign. However, when confined in a layer with the thickness $d$ and prescribed anchoring, the boundary conditions cause $\Lambda$ to have discrete values $\approx d/N$, where $N$ is a positive Integer. So instead of a continuous growth, the modulation period exhibits jumps. In between the jumps, the modulation amplitude decreases to accommodate the smaller splay curvature and with it the effective birefringence increases (Figure 8d). If polar boundary conditions are considered, one jump in birefringence would correspond to a change of $\Lambda$ from $d/N$ to $d/(N-1)$. That is, in a 5 μm cell, $\Lambda$ would increase from the initial $\approx d/N_0$ at the temperature when the first jump occurs to $d/3$ at the temperature where the in-plane modulation becomes visible. For the observed upper limit of 15 jumps, this gives an estimate for the initial $\Lambda$ of $\approx \frac{d}{(15+3)} \approx 280$ nm before the birefringence steps are observed. Alternatively, if no polar orientational anchoring is considered for the N$_S$ phase in the surface of the rubbed cells, one jump could correspond to a change of $\Lambda/2$, i.e., from $2d/N$ to $2d/(N-1)$, in which case the initial modulation period would be $\approx 550$ nm. The modulation period in both cases is large enough to give a weak SHG signal. As observed by POM, the system adapts to an increase of the in-plane modulation period by the edge dislocations, so, likely, the lines of defects at the jumps are also edge dislocations (Figure 8e), in which case a jump would correspond to the change of one period, i.e., from $d/N$ to $d/(N-1)$.

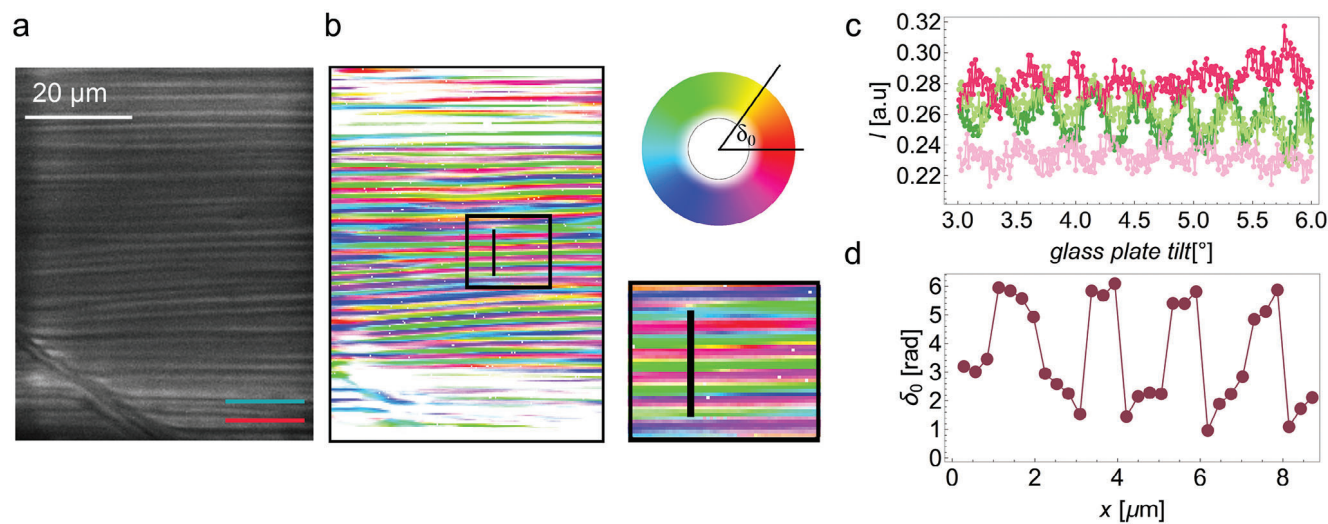

**Figure 7.** SHG interferometry microscopy of FNLC-1571 + 0.6 wt.% of [BMIM][PF$_6$] in a 5 μm parallel rubbed cell at room temperature. a) SHG-I image of the periodic structure observed at room temperature as shown in Figure 5a. Red and green bars indicate the polarization of the incoming IR and reference beam respectively. b) Map across the studied area of the SHG signal phase with respect to the reference SHG beam (see Experimental Section). The phase is color-coded as shown by the color wheel from 0 to 2π. In the regions shown in white, the signal was too low to be determined reliably. The panel includes the enlarged area marked by the black rectangle. c) Examples of SHG interferograms obtained for the regions highlighted by the black rectangle. d) The SHG signal phase across the profile marked by the black line in (b).

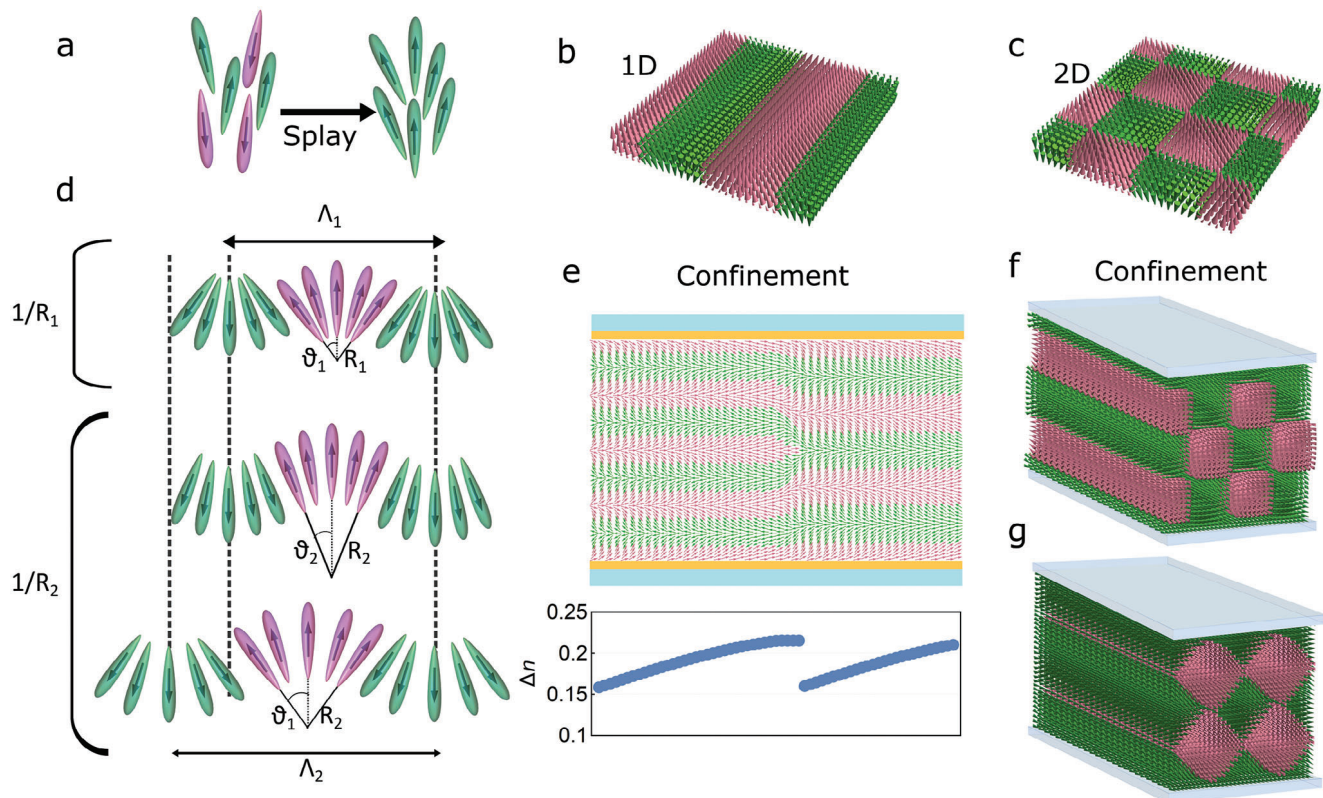

**Figure 8.** a) Schematic representation of flexoelectric effect in nematic liquid crystals. b,c) Schematic representation of 1D and 2D splay nematic phase. d) Schematics of periodic splayed structure characterized by the splay amplitude $\theta$ and modulation period $\Lambda$. The decrease of the splay curvature from $1/R_1$ to $1/R_2$ can be accommodated by either a decrease of $\theta$ (middle row) or an increase of $\Lambda$ (bottom row). e) Schematics of an edge dislocation arising due to a decrease of the splay curvature in a layer, corresponding to the change in one period. Such edge dislocation can be detected in POM as a jump in effective birefringence (bottom graph, calculated considering $n_0 = 1.52$, $\Delta n = 0.22$, and an out-op-plane maximum splay angle varying between 40 deg and 10 deg, jumping back to 40 degrees at the edge dislocation to start decreasing again. f,g) Schematics of two possible orientations of 2D splay modulated structure confined to a layer, with the structure adapting to the prescribed alignment at the surfaces.

The visible in-plane splay modulation together with the observed jumps, which indicate splay modulation in the out-of-plane direction, demonstrate that the structure is periodic in 2D, i.e., consisting of a periodic double splay deformation (Figure 8f,g). The question is whether the antiferroelectric phase in the whole temperature range exhibits 2D periodicity. Insight into this question can be found in the second temperature regime, where jumps in birefringence are observed but in-plane modulation is not visible (Figure 2b,c). The observed structure is full of defects, which is not expected if it would have a 1D out-of-plane modulation. Therefore, it is highly likely that the structure in this region is also 2D modulated with a period too small to be optically visible.

In the first temperature regime just below the N-$N_s$ transition, the POM texture of the pure materials and their mixtures with the ionic liquid look very similar. Whether the structure in this region and in the pure materials also exhibits 2D and not 1D modulation as proposed for some cases[25,27] cannot be ascertained at this stage. The model[32] predicts that the splay nematic phase with 1D modulation is energetically more favorable than the 2D modulated phase only in a narrow temperature range for certain values of the material parameters and that the transition between the 1D to the 2D structure is of the first order. From the cooling experiments under POM, there is no indication that there is a first-order phase transition present. The answer to this question also relates to whether the 2D structure can produce similar experimental features as the 1D structure. The nonresonant and resonant SAXS are the only techniques employed so far that undoubtedly show periodic structure.[23,27] The origin of the SAXS is the modulation of the electron density, which is proportional to the material density. It has been shown that the ferroelectric ordering increases the density of the material,[37,40] which means that a modulation in the magnitude of the polarization will result in a modulation in electron density. The measured SAXS intensity is proportional to the square of the Fourier transform of the density $\langle|\Delta\tilde{\rho}(q)|^2\rangle$. In Figure S11 (Supporting Information), this quantity is calculated for a 1D and a 2D modulated splay structure assuming the variation of the density is proportional to the square of the polarization magnitude $P^2$. In both cases, the first peak appears at $q\Lambda = 4\pi$. In the case of 1D modulation, the second peak is located at $q\Lambda = 8\pi$, while for a square 2D modulated phase besides the second order peak at $q\Lambda = 8\pi$, corresponding to Miller indices (20) and (02), also a weak peak (11) at $q\Lambda = 4\pi\sqrt{2}$ is expected (Figure S11, Supporting Information). In the reported SAXS experiment in the material DIO, only one weak scattering peak was observed and that does not exclude the possibility of the 2D structure.

Finally, to assess the influence of the depolarization field and ions on the splay nematic phase structure, we used a simple 1D model. Similar to what was done in the description of the patterned structures in the $N_F$ phase,[35] the electrostatic terms were added to the free energy density used for the description of the pretransitional behavior at the N–$N_S$ transition.[18,19] In this model, the following assumptions are made: i) $S$ is constant, so the nematic order can be described only by $\mathbf{n}$; ii) $\mathbf{P} = \mathbf{P}_s + \epsilon_0(\epsilon - \mathbf{I})\mathbf{E}$, where $\mathbf{P}_s = P(x)\mathbf{n}$; iii) $\mathbf{n}$ lays in the $xz$ plane $\mathbf{n} = (n_x(x), 0, \sqrt{1-n_x(x)^2})$, and iv) the dielectric tensor is isotropic, $\epsilon = \epsilon\mathbf{I}$. The relevant terms in the free energy can then be written:

$$F = \int \left( \frac{1}{2} K_1 (\mathbf{n}\nabla \cdot \mathbf{n} - \mathbf{S}_0)^2 + \frac{1}{2} K_3 |\mathbf{n} \times (\nabla \times \mathbf{n})|^2 + \frac{1}{4} B(P^2 - P_0^2)^2 \right.$$
$$\left. + \frac{1}{2} K_P (\nabla P)^2 + \frac{1}{2} (\rho_f - \nabla \cdot \mathbf{P}) \phi \right) dV \qquad (1)$$

here, $K_i$ ($i = 1,3$) are the splay and bend elastic constants, and $\gamma$ the flexoelectric coefficient. The flexoelectric term $-\gamma(\mathbf{n} \cdot \mathbf{P})\nabla \cdot \mathbf{n}$ is included in the first term, where $\mathbf{S}_0 = \gamma \mathbf{P}/K_1$ is the ideal splay curvature, which would minimize the splay elastic energy. The sign of $\mathbf{S}_0$ determines the preferred direction of $\mathbf{P}$ when splay deformation is present in the system. The ideal splay curvature that would minimize the splay elastic energy, is $S_0 = \mathbf{n} \cdot \mathbf{S}_0 = \gamma P/K_1$. The first term is larger than the sum of the splay elastic and flexoelectric terms by $\frac{1}{2} K_1 |\mathbf{S}_0|^2$, which is subtracted in the third term. The latter combines the Landau quadratic and quartic terms $\frac{1}{2} AP^2 + \frac{1}{4} BP^4$, with $A$ and $B$ being the corresponding coefficients and $P_0^2 = (\frac{\gamma^2}{K_1} - A)/B$. The fourth term is the first term allowed by symmetry in $\nabla P$, and the last term is the electrostatic term with the charge density of free ions $\rho_f = \rho^+ - \rho^-$, bound charge density $\rho_b = -\nabla \cdot \mathbf{P}$, and $\phi$ the electrostatic potential. Assuming that the free charge density follows the Boltzmann distribution $\rho^\pm = \pm \rho_0 \text{Exp}(\mp e\Phi/(k_BT))$, and that positive and negative ions carry a charge $e = \pm Ze_0$, (where $e_0$ is the elementary charge, $Z$ positive integer, and the charge density $\rho_0$ is the same for positive and negative free ions), then $\rho_f = -2\rho_0 \sinh(e\phi/(k_BT))$. If $|e\phi/(k_BT)| < 1$, then $\rho_f = -\frac{2e^2\rho_0}{k_BT}\phi$, and $\phi$ can be calculated using the linearized Poisson–Boltzmann equation $\nabla^2 \phi = \kappa_D^2 \phi + \frac{1}{\epsilon\epsilon_0} \nabla \cdot \mathbf{P}$, where $\kappa_D^{-1}$ is the Debye screening length $\kappa_D^{-1} = \sqrt{\frac{\epsilon\epsilon_0 k_BT}{2 e\rho_0}}$. We further assumed that the modulated splay structure consists of domains with a constant magnitude of polarization $P_{max}$ and a constant splay deformation, i.e., constant splay curvature $S = s_d S_0$, with neighboring domains having opposite signs of polarization. The domains are separated by Ising domain walls with thickness $d_w$, in which polarization magnitude linearly changes from $+P_{max}$ to $-P_{max}$ or vice versa, and $n_x(x) = \pm(n_{x0} - k(x - x_{wc})^2)$, where $x_{wc}$ is the position of the domain center, $n_{x0}$ the splay amplitude and the coefficient $k$ depends on $\Lambda$, $d_w$, and $s_d$ (**Figure 9a**). It is also assumed that the sample is much larger than the Debye length so that the surface of the sample doesn't affect the structure in the bulk.

For such a structure, the electrostatic potential can be analytically calculated (Note S1, Supporting Information). The free energy Equation (1) is then a function of $\Lambda$, $s_d$, $d_w$, and $P_{max}$. Typically, in realistic cases, the 3rd term in Equation (1) is larger than elastic and flexoelectric terms, and, consequently, $P_{max} \approx P_0$. This term favors small $d_w$. The first term is minimized for $s_d = 1$, i.e., the splay curvature equals $S_0$. The second, i.e., the bend elastic term becomes relevant when $n_{x0}$ is approaching 1. So, the main effect of this term is suppressing the splay amplitude. The fifth, i.e., the polarization gradient term differs from 0 only in the domain wall, and favors large $d_w$. The electrostatic terms favors small $s_d$ and large $d_w$. The constraint for the splay amplitude, $n_{x0} \leq 1$, leads to the constrain for the modulation period $\Lambda \leq \frac{4}{S_0 s_d} + d_w$, which shows that a decrease of the splay curvature may result in an increase of $\Lambda$.

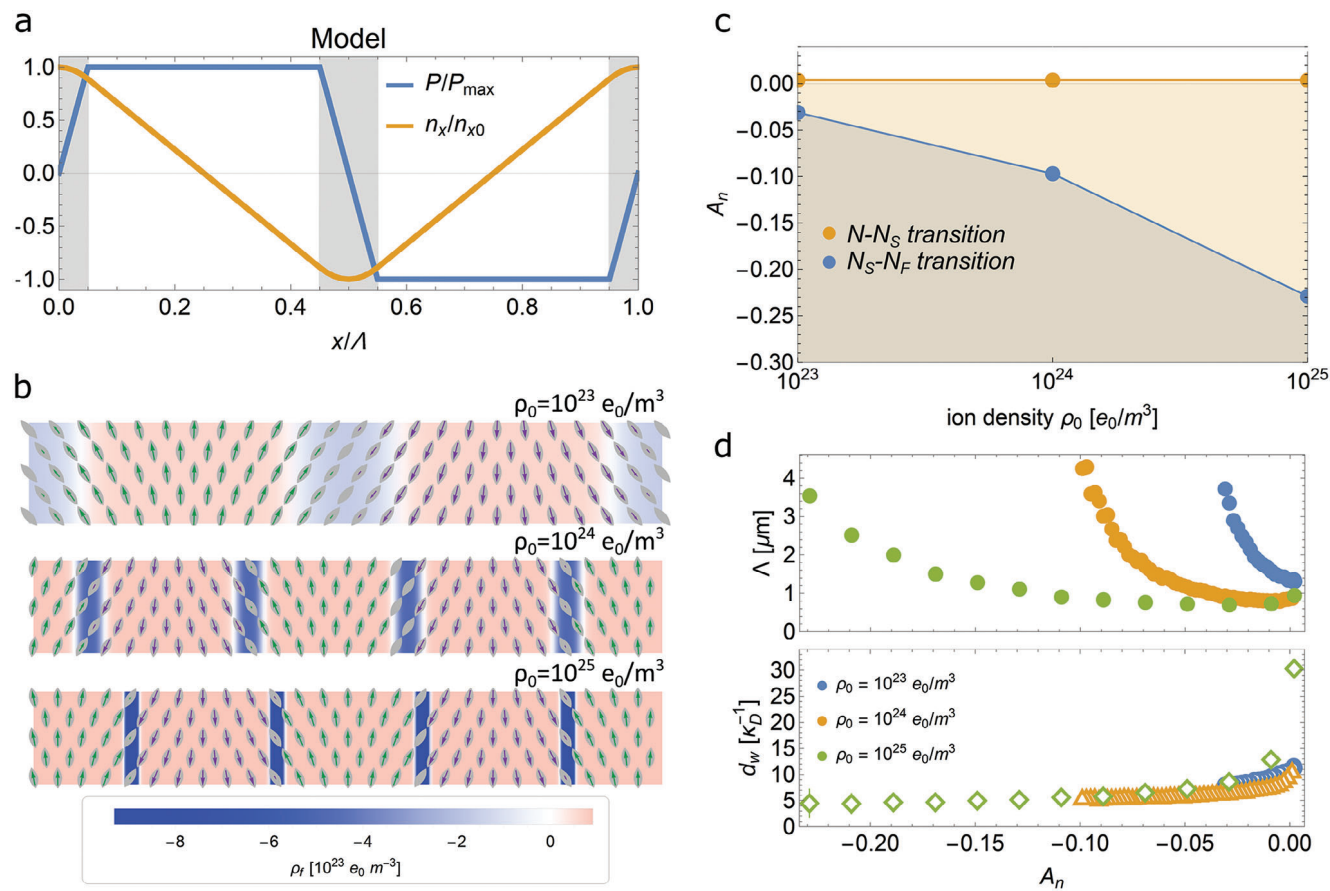

**Figure 9.** a) Polarization and director profile for the 1D modulated splay structure considered in the simple model. b) Representation of the structures obtained by numerical calculations for 3 different ion densities ($10^{23}$, $10^{24}$, and $10^{25}$ $e_0 m^{-3}$) for $A_n = -0.009$, where blue-red colors indicate the concentration of negative (blue) and positive (red) ions. Such distribution screens the bound charges, causing the decrease of the electrostatic potential, and with it the periodicity $\Lambda$ and the ratio between the domain wall thickness ($d_w$) and the periodicity $d_w/\Lambda$ with increasing ion concentration. c) Phase diagram showing the stability of the $N_S$ phase as a function of $A_n$ and $\rho_0$, assessed by comparing the free energy density averaged over a period to the free energy density of the uniform $N_F$ phase. d) Calculated dependence of the $\Lambda$ and $d_w$ on the parameter $A_n$. The Debye length $\kappa_D^{-1}$ is 26, 8.3, and 2.6 nm for $10^{23}$, $10^{24}$, and $10^{25}$ $e_0 m^{-3}$, respectively.

The numerical minimization of the free energy was performed for the following set of parameters: $K_1 = K_3 = 20$ pN, $\gamma = -0.01$ V, $A = \frac{A_n}{(\varepsilon \varepsilon_0)}$, $B = 4.5 \cdot 10^9$ (Nm$^6$) (As)$^{-4}$, $K_P = 8 \cdot 10^{-10}$ Nm$^4$ (As)$^{-2}$, $\varepsilon = 100$, and $Z = 1$ (Note S1, Supporting Information). Here, the parameter $A_n$ plays the role of the temperature, and parameter $B$ is chosen so that at $A_n = 1$, $P_0 = 0.05$ As m$^{-2}$. In Figure 9b, the structures for 3 different ion densities ($10^{23}$, $10^{24}$, and $10^{25}$ $e_0 m^{-3}$) are compared for the value of $A_n = -0.009$. The free ions redistribute so that in the domain walls, there is a larger concentration of negative ions (blue) while in the domains the positive ones (red) dominate. These charges screen the bound charges and cause a decrease of the electrostatic potential. Both, $\Lambda$ and the ratio $d_w/\Lambda$ decrease with increasing ion concentration. This is a consequence of the suppression of the electrostatic term, which favors large $d_w$ and small $s_d$, with the latter also promoting larger $\Lambda$. Figure 9c shows the stability of the $N_S$ phase as a function of $A_n$ and $\rho_0$. The stability of the modeled structure for $N_S$ phase was determined by comparing the free energy density averaged over $\Lambda$ to the free energy density of the uniform structure of $N_F$ phase. Similar to what is observed in experiments, the addition of ions significantly increases the stability of the $N_S$ phase. In Figure 9d, the dependence of $\Lambda$ and $d_w$ on the parameter $A_n$ is shown. For all studied ion densities away from the N–$N_S$ transition, $\Lambda$ increases, while $d_w$ decreases with the decreasing $A_n$ to ≈5 Debye lengths $\kappa_D^{-1}$.

This behavior of $\Lambda$ is in qualitative agreement with the experimental observations (Figure 3c) and it can be understood as follows. With decreasing $A_n$, $P_{max}$ grows, and, consequently, the electrostatic energy increases that as discussed before results in a larger $\Lambda$. The behavior of $d_w$ cannot be compared with the experiments, because the Ising domain walls are not seen by POM, and due to limited spatial resolution they were also not resolved in SHG microscopy. The dependence of $d_w$ on $A_n$ is a consequence of the competition between the electrostatic and the 3rd, i.e., the polarization term in Equation (1) that both increase with $P_{max}$. While the electrostatic term favors a larger $d_w$, the polarization term suppresses it, and because the latter increases faster with $P_{max}$ it prevails so that $d_w$ decreases with the decreasing $A_n$.

The influence of ions on the $N_S$ phase has been also studied theoretically by Paik and Selinger.[41] In their approach, the ions were included in the free energy indirectly through the screening length, and the spatial dependence of **n** and **P** were modeled with sine and cosine waves in 1D. The sinusoidal form is good close to N–$N_s$ transition while at lower temperatures, the previous numerical calculations[18,32] show that the shape is closer to our model. In this sense, the approaches presented here and in ref. [41] are complementary. Both models demonstrate competing roles of electrostatics and flexoelectric terms in the free energy, the first favoring the uniform $N_F$ phase while the second modulated $N_s$ phase. The advantage of the model presented here is that it predicts the temperature dependence of Λ that is in qualitative agreement with the experimental observation while its drawback is that the free energy minimization does not give analytical solutions. The model in ref. [41] is solved analytically and predicts conditions for the N–Ns transition and the behavior of Λ close to this phase transition, however, it does not predict correctly the behavior of Λ deeper in the $N_S$ phase because the sinusoidal form of **n** and **P** is not a good approximation there.

## 4. Conclusion

In conclusion, we have shown that the addition of ionic liquid enhances the stability of the antiferroelectric nematic phase in RM734 and mixture FNLC-1571. POM, SHG-M, and SHG-I experiments reveal a 2D modulated antiferroelectric splayed structure with modulation periods up to 10 μm, demonstrating that the flexoelectric coupling is the driving mechanism for the appearance of the antiferroelectric nematic phase. These findings support the prediction that the antiferroelectric phase is indeed a splay nematic phase $N_s$.[18,32] To distinguish between the 1D and 2D modulated cases, we propose the following notation: $N_{S1}$ for 1D and $N_{S2}$ for 2D structure, while $N_S$ can be used in cases where the dimensionality is not resolved. From the present analysis, the question of whether the 2D splay nematic phase is more stable than the 1D splay nematic phase in all materials and/or in the whole temperature range remains open. This issue is also closely linked to the challenge of resolving the dimensionality at modulation periods that extend beyond visible but are too large to be detected in SAXS experiments. Additionally, the observed nematic-like superstructure, with its rich variety of topological defects formed by the modulated structure, raises new questions about the topology, geometry, and physics of these defects.

## 5. Experimental Section

*Materials*: Liquid crystalline materials RM734 and FNLC-1571 (Merck Electronics KGaA) were mixed with a small amount (from 0.0005 to 5 wt.%) of ionic liquid ([BMIM][PF$_6$], Sigma–Aldrich). The mixtures of RM734 and [BMIM][PF$_6$] were prepared by mixing RM734 with a solution of [BMIM][PF$_6$] in chloroform, followed by solvent evaporation. The mixture of FNLC-1571 and [BMIM][PF6] was prepared by directly mixing the liquids and subsequently heating the mixture in the isotropic phase. The studied concentrations of ionic liquid correspond approximately to ion number densities $\rho_N$ ranging from $10^{22}$ to $10^{26}$ ions m$^{-3}$. The mixtures were filled in LC- cells with 5 and 10 μm thicknesses and surface treatments (EHC planar cells with parallel rubbing – rubbed polyimide promoting in-plane alignment of the director, ITO electrodes without any surface layer, hydrophilic coating nonafluorohexyltriethoxysilane (Gelest)). In the cells with rubbed surfaces, all samples showed good homogeneous alignment in the nematic phase with the director **n** parallel to the rubbing direction. The phase behavior of the mixtures was determined by POM. The phase transition temperatures as determined on cooling (1 K min$^{-1}$) were the following: RM734 (N (128.8 ± 0.5 °C) $N_S$ (127.8 ± 0.5 °C) $N_F$), RM734 + 0.005 wt.% of [BMIM][PF$_6$] (N (128 ± 1 °C) $N_S$ (126.7 ± 1 °C) $N_F$), RM734 + 0.05 wt.% of [BMIM][PF$_6$] (N (127.8 ± 1 °C) $N_S$ (124.7 ± 1 °C) $N_F$), RM734 + 0.5 wt.% of [BMIM][PF$_6$] (N (125.7 ± 1 °C) $N_S$ (111.2 ± 3 °C) $N_F$), RM734 + 5 wt.% of [BMIM][PF$_6$] (N (105.5 ± 1 °C) $N_S$ (phase coexistence 66–73 °C) $X_I$); FNLC-1571 + 0.5 wt.% of [BMIM][PF$_6$] (N (58.2 ± 1 °C) $N_S$ (<5 °C) $N_F$) The transition temperatures determined by POM observations have two sources of error: the first arises from the temperature gradient in the heating stage (0.5 K), and the second from variations in [BMIM][PF$_6$] concentration, causing the transition temperatures to differ between batches (0.5 K for the N–$N_S$ transition and up to 2.5 K for the $N_S$–$N_F$ transition). In RM734, the $N_S$ and $N_F$ phases were monotropic, and they remain monotropic in the mixtures as well. No signs of phase separation were observed in the samples with [BMIM][PF6] concentrations up to 0.6 wt.%. The mixture of RM734 with 5 wt.% [BMIM][PF$_6$] appears homogeneous in the N phase, in the $N_S$ phase within a temperature range of 26 K below the N–NS transition, and in the lower-temperature optically isotropic gel phase. About 26 K below the N–$N_S$ transition, the structure in this mixture becomes grainy. However, it cannot be determined from these observations alone whether this was due to a dense network of defects or indicative of phase separation. At the transition between the $N_S$ phase and the optically isotropic phase, there was a phase coexistence region of ≈7 K, suggesting that this mixture was prone to phase separation under certain conditions.

*Polarizing Optical Microscopy (POM)*: The materials filled in the LC cells that were placed in a heating stage were examined by a polarizing optical microscope (Nikon OptiHot 2 – POL) using SLWD objectives. The images were taken by Canon EOS M200 camera. The temperature gradient within the sample in the heating stage at the cooling rate of 0.1 K min$^{-1}$ was <0.25 K mm$^{-1}$ as was estimated from the motion of the phase front at the N–$N_s$ transition. The temperature behavior of the samples was typically studied between crossed polarizers with the sample rotated so that the rubbing direction was at an angle of ≈20 deg with respect to the polarizer (or analyzer). At given temperatures, the samples were investigated between crossed polarizers, with inserted lambda plate, with analyzer uncrossed for ± 20 deg, and the sample rotated for ± 20 deg to deduce information on a particular structure. To estimate the magnitude of the birefringence jumps, a monochromatic filter (546 nm) and a Sénarmont compensator inserted into the optical train of the microscope at a 45-degree angle were used.

The modulation period of the periodic structure shown in Figure 3c was determined by performing a 1D Fourier analysis of the red channel of the image and calculating the period as $2\pi/q_{min}$, where $q_{min}$ is the position of the first peak in the Fourier transform.

*Second Harmonic Generation Microscopy (SHG-M) and Interferometry (SHG-I)*: Second Harmonic Generation microscopy (SHG-M) and interferometry (SHG-I) were performed using a custom-built scanning microscope with a pump Erbium-doped fiber laser (C-Fiber A 780, MenloSystems, 785 nm, 95 fs pulses at a 100 MHz repetition rate). A detailed description of the setup is available in ref. [35] For the SHG interferometry (SHG-I), a BBO reference crystal was inserted before the sample followed by a Michelson interferometer for time compensation between the reference and the fundamental pulse. The phase of the reference was controlled by the rotation of a glass plate mounted on a motorized rotator. The SHG-M and SHG-I images were acquired with objective and recorded with a high-performance CMOS camera (Grasshopper 3, Teledyne Flir).

In the materials RM734 and FNLC-1571, the d333 process mostly contributes to the SHG signal.[35] This means that the SHG signal was the strongest when laser polarization was parallel to the director. This enabled us to estimate the director's average direction by measuring the SHG signal's dependence on the laser polarization's orientation.

In the SHG interferometry setup, the recorded intensity can be expressed as $I = (A_{sample}\sin(kz) + A_{reference}\sin(kz + \delta))^2$. The phase between the SHG signal of the sample and that of the reference was varied by inserting a glass slide (thickness d) and varying the incidence angle ($\phi$) and then $\delta$ can be expressed as:

$$\delta = \delta_0 + \frac{kd}{\sqrt{1 - \frac{\sin^2(\phi)}{n^2}}} \quad (2)$$

where $k = 2\pi n/\lambda$. To obtain $\delta_0$, SHG-I experiments (Figure 7) were analyzed as follows. Detected intensities were averaged over small regions (1 × 20 pixels) for each glass plate tilt angle. Then, each interferogram was then fitted to

$$I = A_{sample}^2 \left(1 + A \sin\left(\delta_0 + \frac{\frac{2\pi n d}{\lambda}}{\sqrt{1 - \frac{\sin^2(\varphi - \varphi_0)}{n^2}}}\right)\right)^2 \quad (3)$$

where $\lambda = 780/2$ nm, the refractive index is fixed to $n = 1.516$, and the thickness of the glass slide is taken as d = 1.046 mm. $\varphi_0$ was determined by calibrating the initial angle of the glass plate and fixed for all the data sets.

## Supporting Information

Supporting Information is available from the Wiley Online Library or from the author.

## Acknowledgements

The authors wish to acknowledge Simon Čopar for his valuable insights and discussions. The ferroelectric nematic liquid crystal FNLC-1571 used in this work was supplied by Merck Electronics KGaA. P.M.R., E.H., M.L., N.O., N.S., and A.M., acknowledge the support of the Slovenian Research and Innovation Agency (grant numbers P1-0192, J1-50004, N1-0195, and PR-11214). R.J.M. thanks UKRI for funding via a Future Leaders Fellowship, grant no. MR/W006391/1, and the University of Leeds for funding via a University Academic Fellowship.

## Conflict of Interest

The authors declare no conflict of interest.

## Author Contributions

P.M.R. and E.H. contributed equally to this work. A.M. conceived the hypothesis and designed the study. E.H., P.M.R., M.L., N.O., N.S., and A.M. performed the experiments. P.M.R., E.H., N.S., and A.M. collected and analyzed the data. A.M. carried out the model calculations. R.M. and C.J.G. provided professional advice about the materials and the research. A.M. supervised the project. A.M., N.S., and P.M.R. wrote the first draft of the manuscript. All authors read and approved the final manuscript.

## Data Availability Statement

The data that support the findings of this study are available from the corresponding author upon reasonable request.

## Keywords

# ADVANCED SCIENCE
Open Access

## Supporting Information

Antiferroelectric Order in Nematic Liquids: Flexoelectricity Versus Electrostatics

*Peter Medle Rupnik, Ema Hanžel, Matija Lovšin, Natan Osterman, Calum Jordan Gibb, Richard J. Mandle, Nerea Sebastián and Alenka Mertelj\**

# Supporting Information

## Antiferroelectric order in nematic liquids: Flexoelectricity vs electrostatics

Peter Medle Rupnik[1,2], Ema Hanžel[1,2], Matija Lovšin[1,2], Natan Osterman[1,2], Calum J. Gibb[3,4], Richard J. Mandle[3,4], Nerea Sebastián[1], and Alenka Mertelj[1]

1 Jožef Stefan Institute, Ljubljana, Slovenia
2 University of Ljubljana, Faculty of Mathematics and Physics, Ljubljana, Slovenia
3 School of Chemistry, University of Leeds, Leeds, UK
4 School of Physics and Astronomy, University of Leeds, Leeds, UK

Corresponding author: alenka.mertelj@ijs.si

## Contents

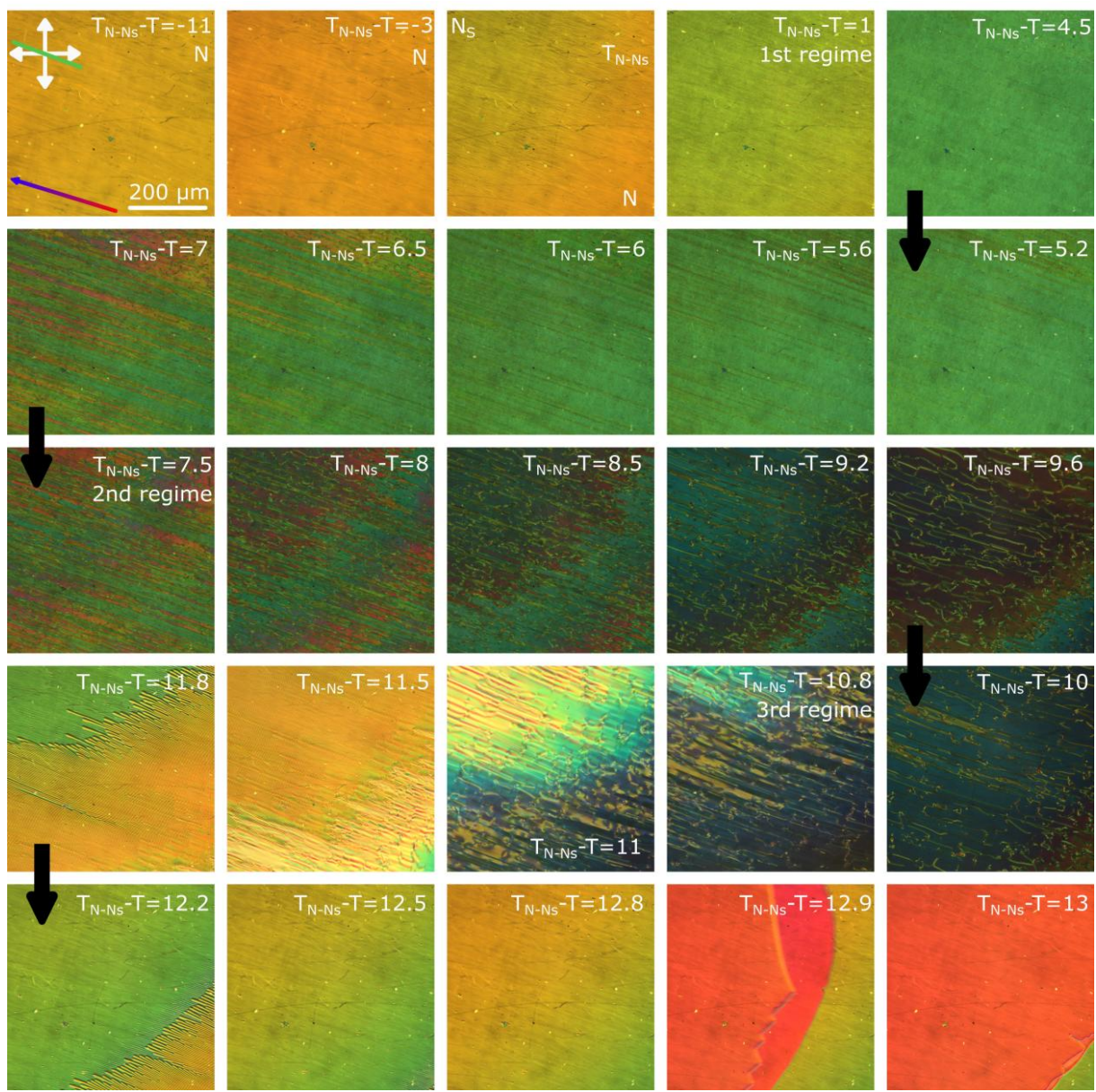

Figure S1. Polarizing optical microscopy images of the textures of RM734 + 0.5 wt% [BMIM][PF$_6$] in a 5μm thick cell with parallel rubbing under a small temperature gradient during a cooling run. Double-headed arrows indicate the direction of the crossed polarizers. The green line indicates the LC cell's rubbing direction. Gradient colored arrow indicates the direction of the temperature gradient, with the blue color (arrowhead) indicating the lower temperature end.

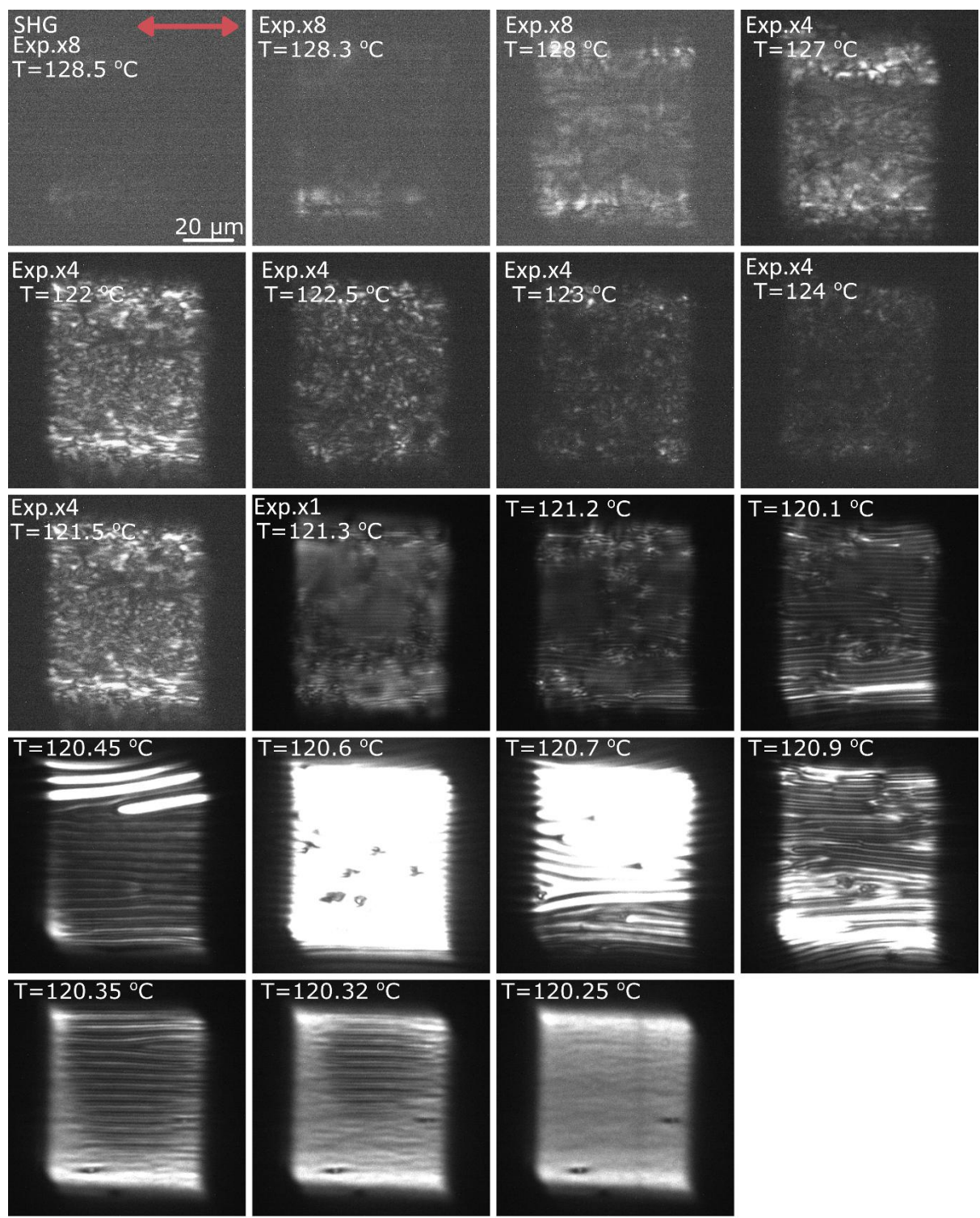

Figure S2. Second Harmonic Generation Microscopy (SHG-M) images of RM734+0.5 wt% [BMIM][PF$_6$] in a 5μm thick cell with parallel rubbing during a cooling run. Double-headed arrow indicates the direction of the incoming light polarization. Images were taken without analyzer. The rubbing direction of the cell is horizontal in the images.

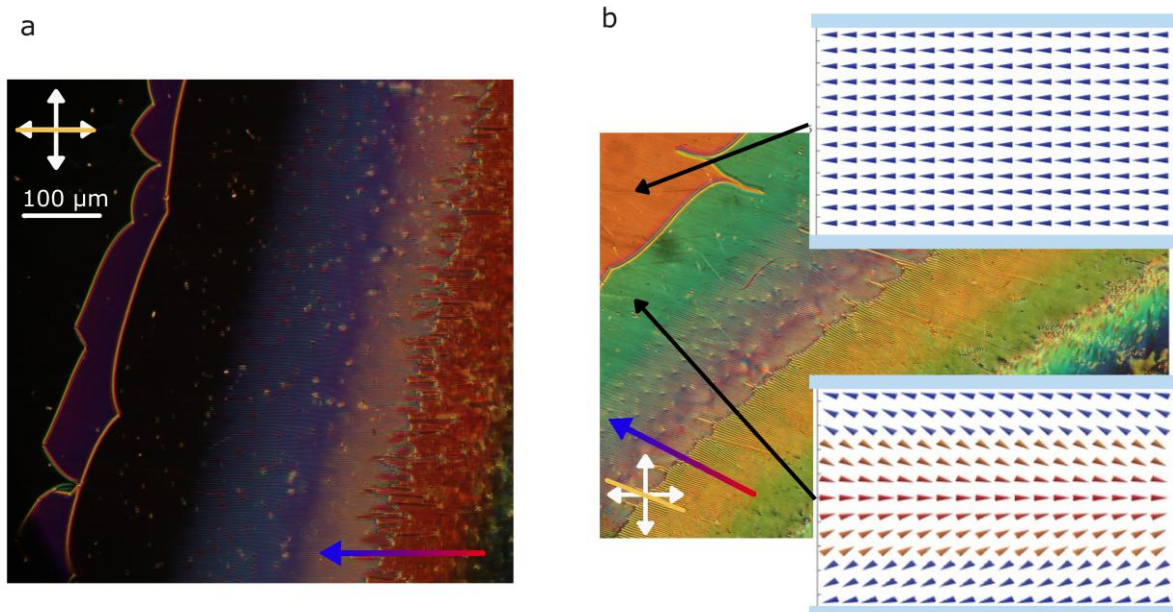

Figure S3. Polarizing optical microscopy images of the textures of RM734+0.5 wt% [BMIM][PF$_6$]. (a) At 110 °C, in a 5 µm thick cell with parallel rubbing highlighting the twisted structure that emerges between the two defect lines dividing homogeneous regions when not located together. (b) Such lines divide two homogeneous regions. This reveals different effective birefringence when the sample is rotated between crossed polarizers. This can be explained by the partial out-of-plane orientation of **n** due to a splay structure, as depicted in the image. Light blue rectangles represent the LC cell surfaces. The estimated difference in effective birefringence between both structures is 0.04, with lower temperature structure having larger value.

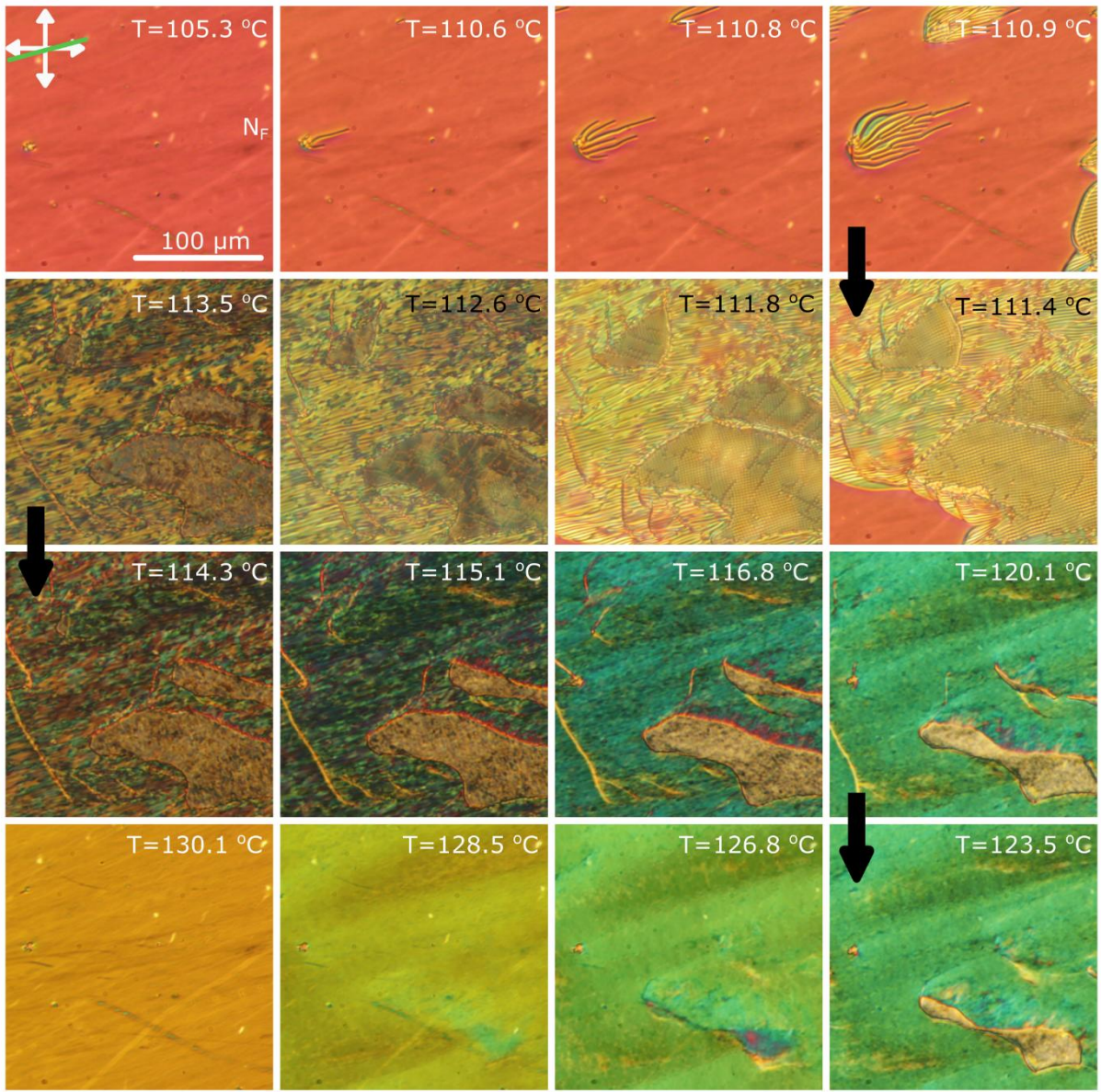

Figure S4. Polarizing optical microscopy images of the textures of RM734+0.5 wt% [BMIM][PF$_6$] in a 10 μm thick cell with parallel rubbing in a heating run (1 °C/min). Double-headed arrows indicate the direction of the crossed polarizers. The green line indicates the LC cell's rubbing direction.

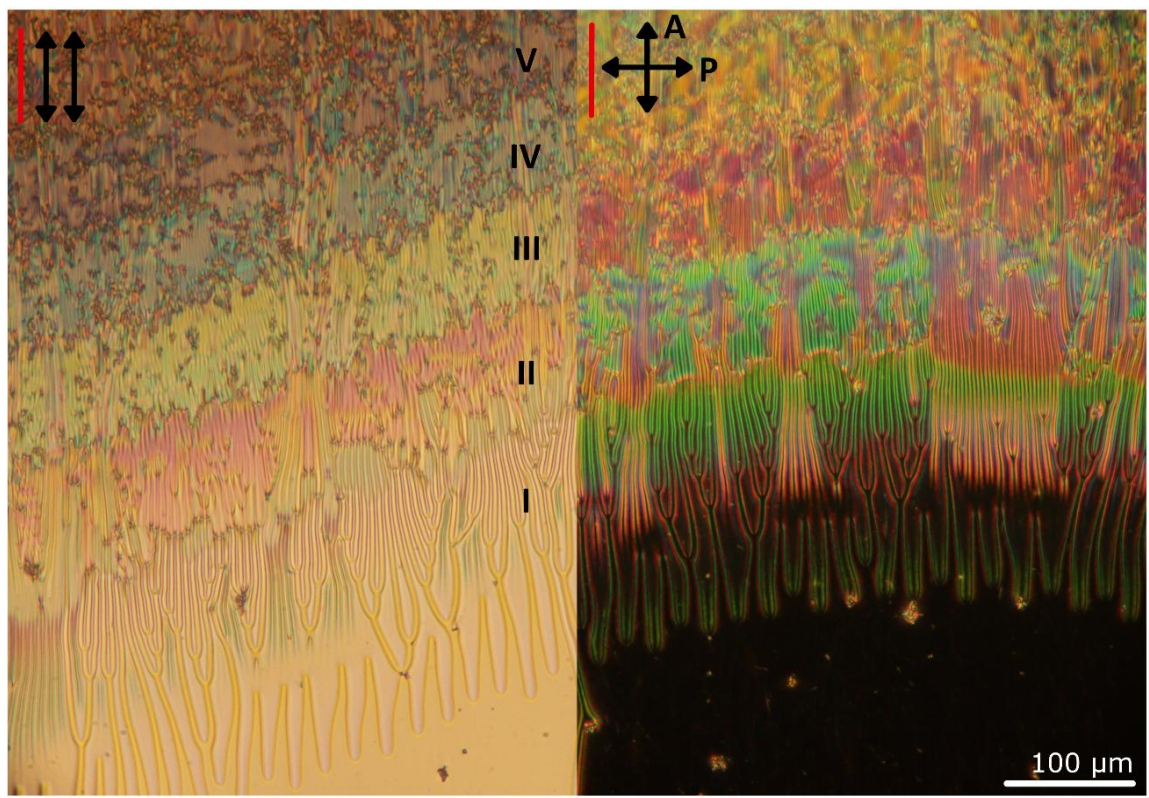

Figure S5. Polarizing optical microscopy images of the textures of RM734 + 0.5 wt% [BMIM][PF$_6$] in a 10 μm thick cell with parallel rubbing at 111 °C. Image is composed of two microphotograph between polarizers at different orientations, indicated by the double-headed arrows; parallel and parallel to the rubbing in the left and crossed in the right. The red line indicates the LC cell's rubbing direction.

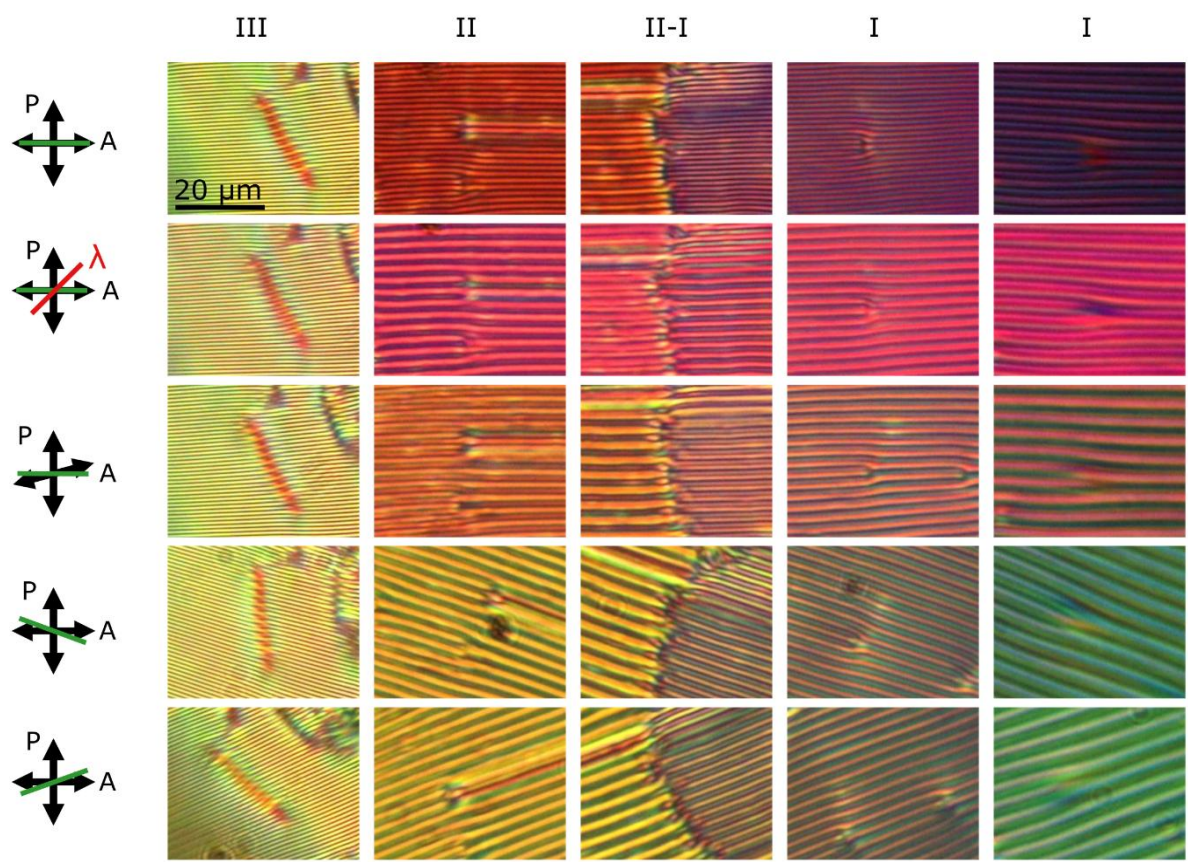

Figure S6. Zoom-in images from the highlighted areas in Figure 3 in the main manuscript at different conditions for for RM734 + 0.5 wt% [BMIM][PF$_6$] in 5 μm parallel rubbed cell. From top to bottom: under crossed polarizers with the cell rubbing direction aligned with the anal, with a full-wave plate inserted at 45 degrees, with analyzer uncrossed by 20 degrees, and with the sample rotated clock- and anticlockwise. Each column corresponds to the different discerned periodic structures, where **I** corresponds to the last one before homogenization.

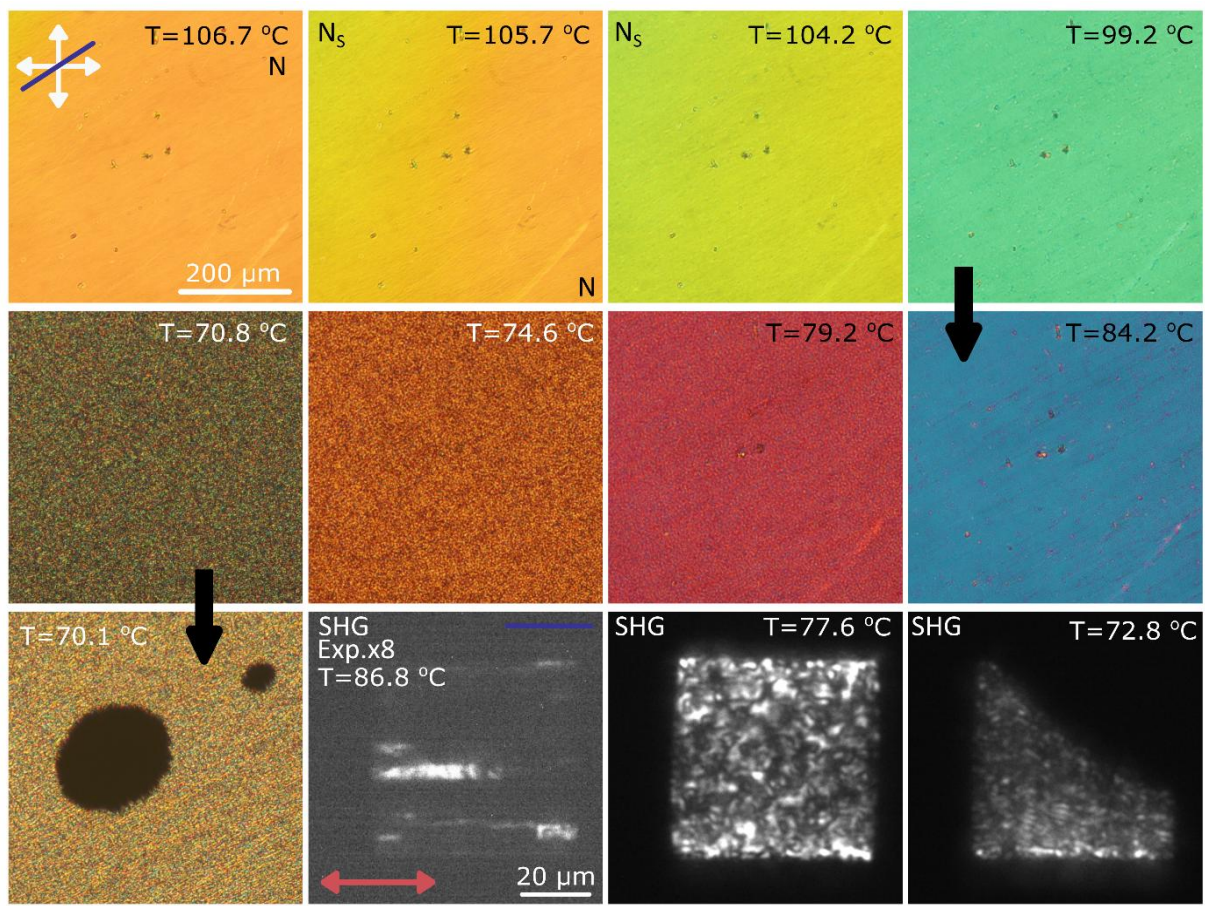

Figure S7. Polarizing optical microscopy images of the textures of RM734 + 5 wt% [BMIM][PF$_6$] in a 5 µm thick cell under cooling at 1 °C/min. 35 K below N-Ns transition the sample exhibits transition to a phase that is optically isotropic. Double-headed arrows indicate the direction of the crossed polarizers. The blue line indicates the LC cell's rubbing direction. In the bottom row, the last three images correspond to SHG-M observations at different temperatures. The sample becomes weakly visible in SHG-M at the N-Ns phase transition and remains so until a grainy structure appears. There the intensity of the M-SHG images becomes stronger and with further cooling decreases. When the material enters the optically isotropic phase, it completely disappears.

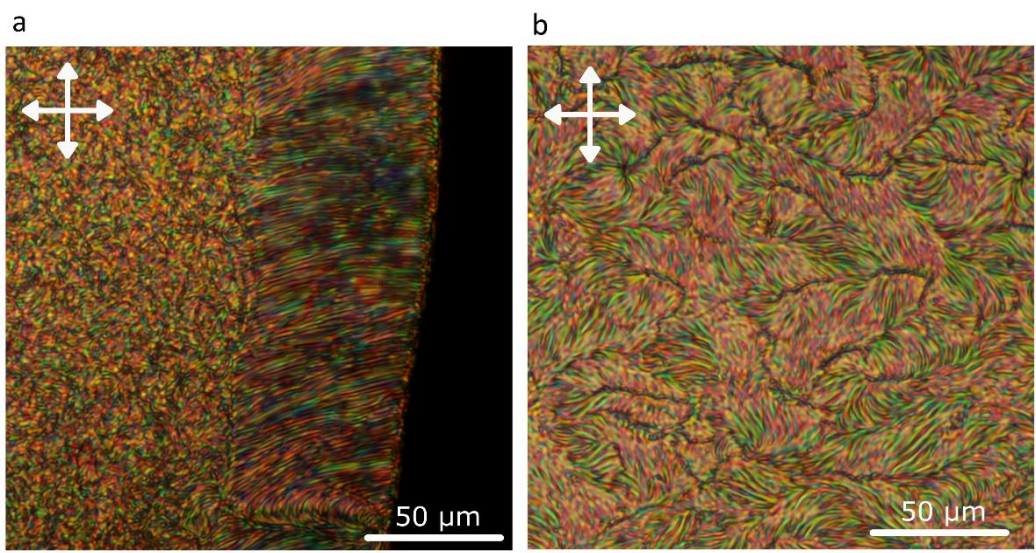

Figure S8. Polarizing optical microscopy images of the textures of FNLC-1571 + 0.6 wt% [BMIM][PF$_6$] in a 10 µm cell with surfaces coated with nonafluorohexyltriethoxysilane. Double-headed arrows indicated the direction of the crossed polarizers.

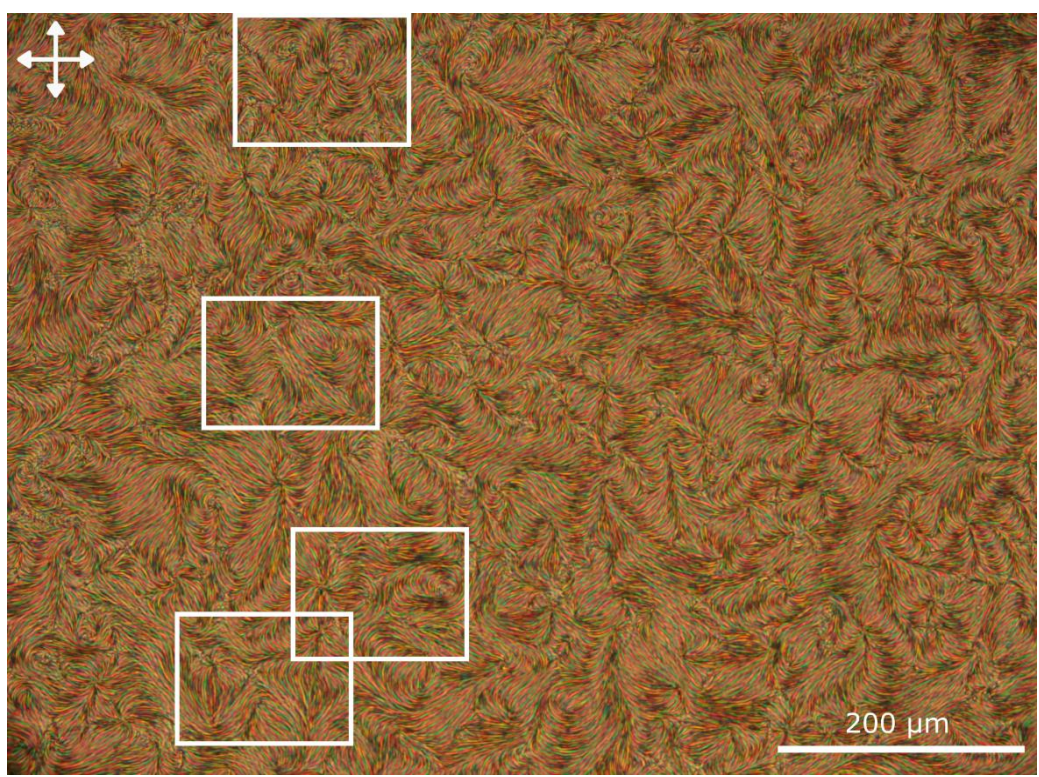

Figure S9. Polarizing optical microscopy textures observed for FNLC-1571 + 0.6wt% of [BMIM][PF$_6$] in a 10 µm LC cell without aligning layers at room temperature outside the ITO electrode area. Double-headed arrows indicate the direction of the crossed polarizers. Rectangles highlight the areas shown in Figure 6 in the main manuscript.

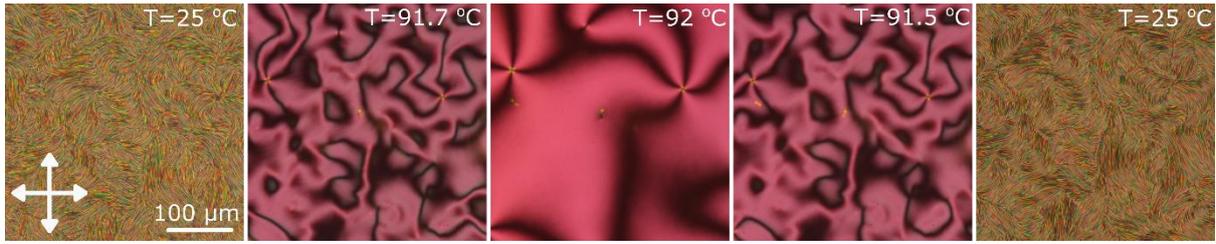

Figure S10. Polarizing optical microscopy textures observed for FNLC-1571 + 0.6wt% of [BMIM][PF$_6$] in a 10 μm LC cell without aligning layers outside the ITO electrode area upon heating from room temperature and subsequent cooling at 0.5 °C/min. Double-headed arrows indicate the direction of the crossed polarizers.

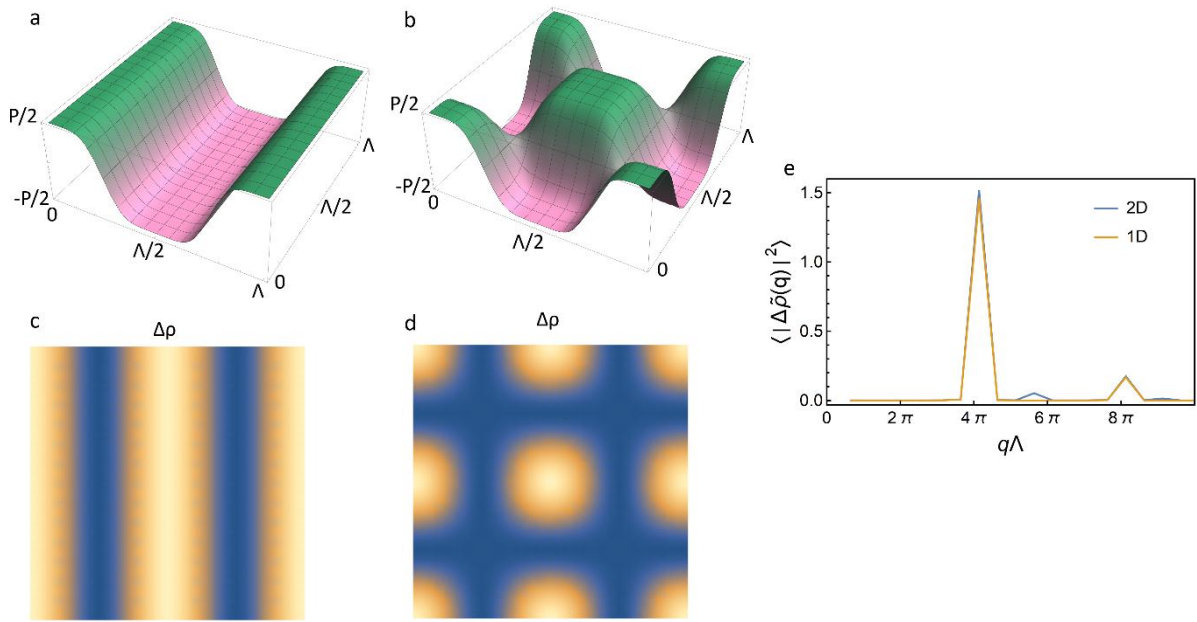

Figure S11. (a-b) 1D and 2D modulation in the polarization magnitude $P$. (c-d) Corresponding modulation in density, proportional to $P^2$, considering that the ferroelectric ordering increases the density of the material [1,2]. (e) Square of the Fourier transform of the density $\langle|\Delta\tilde{\rho}(q)|^2\rangle$ for the structures in (a-d).

## Supplementary Note 1

Similarly as it was done in the description of the patterned structures in the N$_F$ phase [34], the electrostatic terms were added to the free energy density used for the description of the pretransitional behavior at the N-N$_S$ transition [3,4].

In this model, the following assumptions are made: (i) $S$ is constant, so the nematic order can be described only by **n**; (ii) $\mathbf{P} = \mathbf{P}_s + \varepsilon_0(\boldsymbol{\varepsilon} - \mathbf{I})\mathbf{E}$, where $\mathbf{P}_s = P(x)\mathbf{n}$; (iii) **n** lays in the *xz* plane $\mathbf{n} = \left(n_x(x), 0, \sqrt{1 - n_x(x)^2}\right)$, and (iv) the dielectric tensor is isotropic, $\boldsymbol{\varepsilon} = \varepsilon \mathbf{I}$. The relevant terms in the free energy can then be written:

$$F = \int \left(\tfrac{1}{2}K_1(\mathbf{n}\nabla \cdot \mathbf{n} - \mathbf{S}_0)^2 + \tfrac{1}{2}K_3|\mathbf{n} \times (\nabla \times \mathbf{n})|^2 + \tfrac{1}{4}B(P^2 - P_0^2)^2 + \tfrac{1}{2}K_P(\nabla P)^2 + \tfrac{1}{2}(\rho_f - \nabla \cdot \mathbf{P})\phi\right) dV \tag{1}$$

Here, $K_i$ ($i$ = 1,3) are the splay and bend elastic constants and $\gamma$ is the flexoelectric coefficient. The flexoelectric term $-\gamma(\mathbf{n} \cdot \mathbf{P})\nabla \cdot \mathbf{n}$ is included in the first term, where $\mathbf{S}_0 = \gamma \mathbf{P}/K_1$ is the ideal splay curvature, which would minimize the splay elastic energy. The sign of $\mathbf{S}_0$ determines the preferred direction of **P** when splay deformation is present in the system. The ideal splay curvature which would minimize the splay elastic energy, is $S_0 = \mathbf{n} \cdot \mathbf{S}_0 = \gamma P/K_1$. The first term is larger than the sum of the splay elastic and flexoelectric terms by $\tfrac{1}{2}K_1|\mathbf{S}_0|^2$, which is subtracted in the third term. The latter combines the Landau quadratic and quartic terms $\tfrac{1}{2}AP^2 + \tfrac{1}{4}BP^4$, with $A$ and $B$ being the corresponding coefficients and $P_0^2 = \left(\tfrac{\gamma^2}{K_1} - A\right)/B$. The fourth term is the first term allowed by symmetry in $\nabla P$, and the last term is the electrostatic term with $\rho_f = \rho^+ - \rho^-$, and $\phi$ the electrostatic potential. Assuming that the free charge density follows the Boltzmann distribution $\rho^\pm = \pm\rho_0 \text{Exp}(\mp e\Phi/(k_B T))$, and that positive and negative ions carry a charge $e = \pm Z e_0$, (where $e_0$ is the elementary charge, $Z$ positive integer, and the ion density $\rho_0$ is the same for positive and negative free ions), then $\rho_f = -2\rho_0 \sinh(e\Phi/(k_B T))$. If $|e\Phi/(k_B T)| < 1$, then $\rho_f = -\tfrac{2e\rho_0}{k_B T}\phi$, and $\phi$ can be calculated using the linearized Poisson-Boltzmann equation $\nabla^2 \phi = \kappa_D^2 \phi + \tfrac{1}{\varepsilon \varepsilon_0}\nabla \cdot \mathbf{P}$, where $\kappa_D^{-1}$ is the Debye screening length $\kappa_D^{-1} = \sqrt{\tfrac{\varepsilon \varepsilon_0 k_B T}{2 e \rho_0}}$. The bound charges are given by $\rho_b = -\nabla \cdot \mathbf{P}$. We further assumed that the modulated splay structure consists of domains with a constant magnitude of polarization $P_{max}$ and a constant splay deformation, i.e., constant splay curvature $S = s_d S_0$, with neighboring domains having opposite signs of polarization. The domains are separated by Ising domain walls with thickness $d_w$, in which polarization magnitude linearly changes from $+P_{max}$ to $-P_{max}$ or vice versa, and $n_x(x) = \pm(n_{x0} - k(x - x_{wc})^2)$, where $x_{wc}$ is the position of the domain center, $n_{x0}$ the splay amplitude and the coefficient $k$ depends on the $\Lambda$, $d_w$, and $s_d$. It is also assumed that the sample is much larger than the Debye length so that the surface of the sample doesn't affect the structure in the bulk.

For such a structure, the electrostatic potential can be analytically calculated. Using the units of $\kappa_D^{-1}$ for $x$, the solution for normalized electrostatic potential $\phi_n = \tfrac{e\phi}{k_B T}$ in the domain is

$$\phi_{n,d}[x] = -\frac{4\pi\xi_B}{\kappa_D^2}\frac{P_{max}S_0 s_d}{d_w^2}(1 + C_d[d_w, \Lambda]\cosh(x - x_{dc})) = -\frac{P_{max}S_0 s_d}{d_w^2}h_{d,es}[\Lambda, d_w, x] \tag{2}$$

where $C_d[d_w, \Lambda] = \left((24 - \Lambda d_w + 2 d_w^2) \sinh\left(\frac{d_w}{2}\right) - 12 d_w \cosh\left(\frac{d_w}{2}\right)\right) / \left(2 \sinh\left(\frac{\Lambda}{4}\right)\right)$, $x_{dc}$ the position of the domain center, and $\xi_B = \frac{e_0^2}{4\pi\varepsilon\varepsilon_0 k_B T}$ Bjerrum length.

The solution for the domain wall reads:

$$\phi_{n,w}[x] = -\frac{4\pi\xi_B}{\kappa_D^2} \frac{P_{max} S_0 s_d}{d_w^2} \left(12 - \frac{\Lambda d_w}{2} + \frac{d_w^2}{2} + 6(x - x_{wc})^2 + C_d[d_w, \Lambda] \cosh(x - x_{wc})\right) =$$
$$-\frac{P_{max} S_0 s_d}{d_w^2} h_{w,es}[\Lambda, d_w, x]$$

(3)

where $C_w[d_w, \Lambda] = \left((-24 + \Lambda d_w - 2 d_w^2) \sinh\left(\frac{\Lambda - 2d_w}{4}\right) - 12 d_w \cosh\left(\frac{\Lambda - 2d_w}{4}\right)\right) / \left(2 \sinh\left(\frac{\Lambda}{4}\right)\right)$, and $x_{wc}$ the position of the domain wall center.

The free energy (Eq. 1) is then a function of $\Lambda$, $s_d$, $d_w$, and $P_{max}$. In the domains,

$$F_D = A_{yz} \int_{-(\Lambda-2d_w)/2}^{(\Lambda-2d_w)/2} \left(\tfrac{1}{2} K_1 S_0^2 (s_d - 1)^2 + \tfrac{1}{2} K_3 S_0^4 s_d^4 \frac{(x-x_{dc})^2}{1 - S_0^2 s_d^2 (x-x_{dc})^2} + \tfrac{1}{4} B(P_{max}^2 - P_0^2)^2 + f_{d,es}\right) dx$$

(4)

and in the walls:

$$F_w / A_{yz} = \int_{-d_w/2}^{d_w/2} \left(2 K_1 S_0^2 (s_d - 1)^2 \frac{(x-x_{dc})^2}{d_w^2} + 2 K_3 S_0^4 s_d^4 \frac{h_B[x-x_{wc}]}{d_w^2} + \tfrac{1}{4} B \left(P_{max}^2 \frac{4(x-x_{wc})^2}{d_w^2} - P_0^2\right)^2 + \right.$$
$$\left. \tfrac{1}{2} K_P \left(\frac{2 P_{max}}{d_w}\right)^2 + f_{w,es}\right) dx$$

(5)

where

$$f_{d,es} = \frac{P_{max}^2 S_0^2 s_d^2}{d_w^4} \left(-2 e n_0 \, h_{d,es}[\Lambda, d_w, x - x_{dc}]^2 + \frac{k_B T}{2e} d_w^2 h_{d,es}[\Lambda, d_w, x - x_{dc}]\right),$$

$$f_{w,es} = \frac{P_{max}^2 S_0^2 s_d^2}{d_w^4} \left(-2 e n_0 \, h_{w,es}[\Lambda, d_w, x - x_{wc}]^2 + \frac{k_B T}{4e} \left(12(x - x_{w,c})^2 - d_w(\Lambda - d_w)\right) h_{w,es}[\Lambda, d_w, x - x_{wc}]\right),$$

$$h_B[x] = \frac{(4x^3 + x d_w(-\Lambda + d_w))^2}{(16 d_w^2 - S_0^2 s_d^2 (4x^2 + d_w(-\Lambda + d_w))^2)},$$

and $A_{yz}$ is the $yz$ surface of the sample.

The constraint for the splay amplitude, $n_{x0} \leq 1$, leads to the constrain for the modulation period $\Lambda \leq \frac{4}{S_0 s_d} + d_w$.

Typically, in the realistic cases, the 3$^{rd}$ term is larger than elastic and flexoelectric terms, and, consequently, $P_{max} \approx P_0$. This term favours small $d_w$. The first term is minimized for $s_d = 1$. The second, the bend elastic term becomes relevant when $S_0^2 s_d^2 (x - x_{dc})^2$ approaches 1, which is when $n_{x0}$ is approaching 1. So the main effect of this term is suppressing the splay amplitude. The fifth, ie., the polarization gradient term differs from 0 only in the domain wall, and favours large $d_w$. The electrostatic term favours small $s_d$ and large $d_w$.

The numerical minimization of the free energy were performed for the following set of parameters $K_1 = K_3 = 20$ pN, $\gamma = -0.01$ V, $A = \frac{A_n}{(\varepsilon\varepsilon_0)}$, $B = 4.5 \cdot 10^9 \frac{Nm^6}{(As)^4}$, $K_P = 8 \cdot 10^{-10}$ Nm$^4$/(As)$^2$, $\varepsilon = 100$, and $Z = 1$. Here, the parameter $A_n$ plays the role of the temperature, and parameter $B$ is chosen so that at $A_n = 1$, $P_0 = 0.05 \text{ As/m}^2$. The integrand in Eq. 1 was numerically averaged over one modulation period for discrete sets of allowed values of $\Lambda$, $s_d$, and $d_w$ while $P_{max} = P_0$ was fixed. The set with the minimal averaged value of the free energy was chosen as a solution for $\Lambda$, $s_d$, and $d_w$. The minimal averged value was then compared with the free energy density of uniform N$_F$ phase ($n_{x0} = 0$, $P_0 = \sqrt{\frac{-A}{B}}$) to assess the stability of the N$_S$ phase.

Note: When approaching the threshold value of $A_{n,tr}$ at which the transition to N$_F$ happens, the value of the normalized electrostatic potential in the domain walls $\phi_n$ exceeded 1 which means that the linearization of Poisson Boltzmann equation is not that accurate anymore. Comparison of the solution of the linearized Poisson Boltzmann equation with the numerical solution of the full Poisson Boltzmann equation showed that the linearization overestimates $\phi_n$ in the domain walls for up to 20% when approaching $A_{n,tr}$. Consequently, the electrostatic energy is also over estimated, which leads to lower $A_{n,tr}$ and smaller $\Lambda$ compared to those given by solutions of the linearized equation.